\documentclass[%
 reprint,
nofootinbib,
 amsmath,amssymb,
 aps,
 prd,
]{revtex4-1}

\usepackage{graphicx}
\usepackage{dcolumn}
\usepackage{bm}
\usepackage{epsfig}
\usepackage{amsmath}
\usepackage{enumerate}
\usepackage{mathrsfs}
\usepackage{natbib}
\usepackage{hyperref}
\usepackage{hhline}
\usepackage{color}
\usepackage{tcolorbox}
\usepackage{soul}
\definecolor{greyish2}{rgb}{.96,.96,.96}

\newcommand\lsim{\mathrel{\rlap{\lower4pt\hbox{\hskip1pt$\sim$}}
        \raise1pt\hbox{$<$}}}
\newcommand\gsim{\mathrel{\rlap{\lower4pt\hbox{\hskip1pt$\sim$}}
        \raise1pt\hbox{$>$}}}

\begin{document}
\preprint{APS/123-QED}

\title{Construction of Wave Dark Matter Halos: \\Numerical Algorithm and Analytical Constraints}

\author{Tomer D. Yavetz}
\email{t.yavetz@columbia.edu}
\affiliation{Department of Astronomy, Columbia University, New York, NY 10027, USA}
\author{Xinyu Li}
\email{xli@cita.utoronto.ca}
\affiliation{
Canadian Institute for Theoretical Astrophysics, 60 St George St, Toronto, ON, Canada, M5R 2M8\\
Perimeter Institute for Theoretical Physics, 31 Caroline Street North, Waterloo, ON, Canada, N2L 2Y5}
\author{Lam Hui}
 \email{lhui@astro.columbia.edu}
\affiliation{
Center for Theoretical Physics, Department of Physics, Columbia University, New York, NY 10027, USA
}%

\date{\today}

\begin{abstract}
We present a wave generalization of the classic Schwarzschild method for constructing self-consistent halos -- such a halo consists of a suitable superposition of waves instead of particle orbits, chosen to yield a desired mean density profile. As an illustration,  the method is applied to spherically symmetric halos. We derive an analytic relation between the {\it particle} distribution function and the {\it wave} superposition amplitudes, and show how it simplifies in the high energy (WKB) limit. We verify the stability of such constructed halos by numerically evolving the Schr\"odinger-Poisson system. The algorithm provides an efficient and accurate way to simulate the time-dependent halo substructures from wave interference. We use this method to construct halos with a variety of density profiles, all of which have a core from the ground-state wave function, though the core-halo relation need not be the standard one.
\end{abstract}


\maketitle

\section{Introduction} \label{sec:intro}

What makes a halo tick? For a halo built out of particles, be it dark matter or stars, the answer lies in the distribution function $f$, which tells us how the particles are distributed in phase space. A halo in equilibrium should be described by a time-independent $f$, satisfying the Vlasov or collisionless Boltzmann equation (ignoring processes such as two-body encounters). In a classic paper, Schwarzschild \cite{Schwarzschild1979} described a method for self-consistently constructing such a halo with a desired density profile: compute a library of particle orbits in the corresponding gravitational potential, and choose an appropriate weighting of the orbits to reproduce the input density profile, thereby deducing the $f$ that sustains the halo (see also \cite{Richstone1984,BT}). Our goal in this paper is to develop the corresponding method for a halo built out of waves.

Why waves? For a typical halo mass density, such as that of the solar neighborhood, it can be shown that a dark matter candidate with a constituent mass below about $30$ eV behaves like waves, because its de Broglie wavelength falls below the average interparticle separation (\cite{Hui:2021tkt} and references therein). Such a dark matter particle is necessarily bosonic \cite{Tremaine:1979we}, with the prime example being the QCD axion \cite{Peccei:1977hh,Kim:1979if,Weinberg:1977ma,Wilczek:1977pj,Shifman:1979if,Zhitnitsky:1980tq,Dine:1981rt,Preskill:1982cy,Abbott:1982af,Dine:1982ah}, or more generally, an axion-like particle \citep{Svrcek:2006yi,Arvanitaki:2009fg,Hui:2016ltb,Halverson:2017deq,Bachlechner:2018gew}.

Novel wave signatures are most easily observable if the de Broglie wavelength is large, i.e., if the mass is ultra-light ($\lsim 10^{-20}$ eV). This possibility is often referred to as fuzzy dark matter (FDM), a term introduced by Hu, Barkana, and Gruzinov \cite{Hu:2000ke}. There has been a revival of interest in this possibility, starting from the work of Schive, Chiueh, and Broadhurst \cite{Schive:2014dra} \cite{Veltmaat:2016rxo,Schwabe:2016rze,Hui:2016ltb,Mocz:2017wlg,Nori:2018hud,Levkov:2018kau,Bar-Or:2018pxz,Bar:2018acw,Church:2018sro,Li:2018kyk,Marsh:2018zyw,Schive:2019rrw,Mocz:2019pyf,Lancaster:2019mde,Chan:2020exg,Hui:2020hbq} (more generally, the subject of scalar field dark matter has a long history \cite{Baldeschi:1983mq,Turner:1983he,Khlopov:1985jw,Press:1989id,Sin:1992bg,Guzman:1999ft,Matos:1999et,Peebles:2000yy,Goodman:2000tg,Lesgourgues:2002hk,Matos:2003pe,Amendola:2005ad,Chavanis:2011zi,Suarez:2011yf,RindlerDaller:2011kx,Berezhiani:2015bqa,Fan:2016rda,Alexander:2016glq}; see reviews \cite{Niemeyer:2019aqm,Ferreira:2020fam,Hui:2021tkt} and references therein).

The Schwarzschild-inspired strategy for building a halo of a given desired density profile involves computing a library of wave eigenmodes (replacing particle orbits) in the corresponding gravitational potential, and choosing an appropriate superposition of them to reproduce, or match as best as we can, the given density profile. Widrow and Kaiser proposed the {\it wave} superposition amplitude is proportional to the square root of the {\it particle} distribution function $f$ \cite{Widrow:1993qq}. We will demonstrate that this approach is to a large extent correct, with important corrections in the inner parts of a halo. More recently, \citet{Lin:2018whl} constructed wave dark matter halos using the distribution of amplitudes measured from dynamical FDM simulations.

We build on these earlier works with three principal objectives in mind: (1) to clarify the validity and limitation of the Widrow-Kaiser ansatz through a systematic exploration of the particle and wave descriptions in the WKB limit; (2) to extend the \citet{Lin:2018whl} construction to halos of a general density profile,\footnote{Realistic halos have a variety of profiles, due in part to feedback processes. It is thus useful to construct halos with profiles beyond those predicted by pure dark matter cosmological simulations.}
and show that the wave construction inevitably puts constraints on what kind of density profile is allowed; and (3) to demonstrate that such a wave halo construction faithfully reproduces wave interference substructures seen in dynamical wave simulations.

The last point is a particularly important motivation for our work: dynamical wave simulations that evolve the exact Schr\"odinger-Poisson system are computationally expensive. The halo construction method \`a la Schwarzschild, though numerical, is considerably more efficient than these dynamical wave simulations. This opens the door to detailed investigations of wave halo substructure. Indeed, there are a few recent papers heading in this direction: \citet{Dalal:2020mjw} employed the method of Widrow and Kaiser to construct wave dark matter halos, for the purpose of studying the scattering of tidal streams by the interference substructure; we in \cite{Li:2020ryg} studied soliton random walk and oscillations by decomposing a wave dark matter halo into its eigenmodes; \citet{Zagorac:2021qxq} studied the distortions of solitons using perturbation theory.

The outline of the paper is as follows. We present the wave halo construction method in \S \ref{sec:method}, focusing on the particular example of a Navarro-Frenk-White (NFW) \cite{Navarro:1996gj} halo with a core according to \citet{Schive:2014hza}. We use this example to discuss several choices one may make in such a construction. In \S \ref{sec:sims}, we check the stability of the so-constructed halo by performing dynamical wave simulations. We then turn to the construction of halos with more general density profiles in \S \ref{sec:profiles} -- as we will see, a cored central profile is a universal feature of wave dark matter, while a variety of outer profiles is allowed. We explore in \S \ref{sec:core-halo} halos with a core size that deviates from the \citet{Schive:2014hza} halo-core relation, and conclude in \S \ref{sec:disc}. Appendix \ref{sec:app} contains a derivation of the connection between the particle description (in terms of distribution function) and the wave description (in terms of superposition amplitudes), in particular verifying the Widrow-Kaiser ansatz in the WKB limit.

Throughout this paper, we adopt a dark matter mass of the ultra-light kind for illustration, but it should be emphasized that the method can be applied in principle to wave dark matter of any mass ($\lsim 30$ eV). While the ultra-light end of the spectrum is the most interesting from the point of view of astrophysical observations, axion detection experiments target a wide range of axion masses -- from $10^{-22}$ to $10^{-3}$ eV -- and are sensitive to wave interference features down to very small scales.

We also use the terms fuzzy dark matter (FDM) and wave dark matter somewhat interchangeably, even though FDM is more appropriately restricted to wave dark matter of the ultra-light kind \cite{Hui:2021tkt}. 

One more remark on terminology -- in the discussion below we utilize (and draw a contrast between) two different kinds of numerical simulations:
one is the Schwarzschild-like construction of halos which allows one to efficiently simulate and evolve the halo substructure; the other is the computationally expensive simulation that evolves the Schr\"odinger-Poisson system exactly.
We refer to the former as \textit{Schwarzschild simulations} and to the latter as {\it dynamical wave simulations}.

\section{Method for Wave DM Halo Construction} \label{sec:method}

The behavior of wave dark matter is described by a complex scalar field $\Psi(\bm{x},t)$ that obeys the Schr{\"o}dinger-Poisson (SP) equations:
\begin{equation} \label{eq:schrod-pois1}
    i\hbar\frac{\partial\Psi}{\partial t} = \bigg(-\frac{\hbar^2}{2m_a}\nabla^2+m_aV\bigg)\Psi \ ,
\end{equation}
\begin{equation} \label{eq:schrod-pois2}
    \nabla^2 V = 4\pi G\rho = 4\pi Gm_a|\Psi|^2 \ ,
\end{equation}
where $V$ is the gravitational potential and $m_a$ is the particle mass.
We set $m_a = 8.1\times10^{-23}$ eV/c$^2$ throughout this work. We refer to $\Psi$ as the wave function, with $|\Psi|^2$ giving the number density of particles. It is important to note that $\Psi$ is a \textit{classical} complex field, describing the regime in which there are many dark matter particles per de Broglie volume. Quantum fluctuations in this regime are therefore negligible.\footnote{Note that a factor of $\hbar$ still appears in Eq. \ref{eq:schrod-pois1} despite the claim that $\Psi$ is a classical field. However, by dividing Eq. \ref{eq:schrod-pois1} by a factor of the particle mass $m_a$ it is evident that $\hbar$ only appears in conjunction with $m_a$, and one may simply think of the quantity $\hbar / m_a$ as setting a characteristic scale in the problem.}

Our goal is to construct an FDM halo in local equilibrium, whose potential profile is static. Note that a realistic halo, even a virialized one, has short timescale fluctuations.
Here, we are interested in a halo that has no long term evolution, for which as a first approximation, the potential $V$, and therefore the Hamiltonian, can be treated as time-independent. In this case,
$\Psi(\bm{x},t)$ can be decomposed into a series of normalized and orthogonal spatial eigenmodes $\psi_j (\bm{x})$ that satisfy the time-independent Schr{\"o}dinger equation:
\begin{equation} \label{eq:schrod-indep}
    \bigg(-\frac{\hbar^2}{2m_a}\nabla^2 + m_aV\bigg)\psi_j = E_j \psi_j  \ .
\end{equation}

Each eigenmode $\psi_j$ has an associated (time-independent) complex amplitude $a_j$ and frequency $\omega_j$, the latter of which is related directly to that state's energy eigenvalue ($\omega_j = E_j / \hbar$). The time-dependent wave function can thus be written as a sum of the eigenmodes:
\begin{equation} \label{eq:wave-func-full}
    \Psi(\bm{x},t) = \sum_j a_j \psi_j(\bm{x}) e^{-iE_jt/\hbar} \ ,
\end{equation}
and the number density of particles is:
\begin{eqnarray} \label{eq:number-dens}
    \nonumber & & |\Psi(\bm{x},t)|^2 = \bigg|\sum_j a_j \psi_j(\bm{x}) e^{-iE_jt/\hbar} \bigg|^2 \\
    \nonumber & & = \sum_j |a_j|^2 |\psi_j(\bm{x})|^2 + \sum_{j\neq k} a_ja_k^*\psi_j(\bm{x})\psi_k^*(\bm{x})e^{i(E_k - E_j)t/\hbar} \ . \\
    & &
\end{eqnarray}

The final term represents the interference of different eigenstates, and is responsible for the small-scale FDM fluctuations (sometimes referred to as `granules'). The phase of each eigenmode (phase of $a_j$) is assumed random, and thus the granules or interference fringes take on a somewhat random pattern. 
The typical granule size is given by the de Broglie wavelength, $\hbar/(m_a v)$, where $v$ is roughly the velocity dispersion of the halo \citep{Schive:2014dra,Hui:2016ltb, Li:2020ryg}.

The interference term is manifestly time-dependent, with a characteristic timescale of the order of the de Broglie time $\hbar/(m_a v^2)$. In other words, the density, and therefore the gravitational potential, is time-dependent in detail.\footnote{It is worth noting that even a halo composed of classical particles such as stars (i.e., negligible de Broglie wavelength) has time-dependent fluctuations. The analog of random phase for the wave eigenmodes is random phase for the stellar orbits. Occasionally, some stars might come together, creating temporary density enhancements. Such density fluctuations become small if the density of stars is sufficiently high.}
However, under time averaging, or averaging over the random phases:
\begin{equation} \label{eq:wave-func-diag}
    \langle |\Psi(\bm{x},t)|^2 \rangle = \sum_j |a_j|^2 |\psi_j (\bm{x})|^2 \ .
\end{equation}

Our task is to find the superposition coefficients $a_j$ such that
the averaged density profile (Eq. \ref{eq:wave-func-diag}) 
matches, to the extent possible, the desired density profile. Once this is done, the actual density profile at any given moment (Eq. \ref{eq:number-dens}) exhibits the time-dependent halo substructure from wave interference.
The evolution of the substructure is completely determined by this halo construction: it is just a matter of attaching the right time-dependent phase to each eigenmode.

It is important to emphasize, however, that this evolution is approximate -- the exact evolution should account for the fact that the gravitational potential fluctuates with time, i.e., a completely self-consistent treatment involves solving the Schr\"odinger-Poisson system (what we call dynamical simulations). We will carry out dynamical simulations in \S \ref{sec:sims} to demonstrate that our time-independent halo construction method works: that the constructed halo is stable, and that the substructures appear very similar to those seen in dynamical simulations. The advantage of our halo construction is that it is much faster, and evolving the substructure (by propagating the phases of eigenmodes) takes minimal computational effort.

To illustrate the application of this procedure, we impose spherical symmetry throughout the remainder of this work. However, our method is more general: it could be used to generate triaxial halos if one wishes, though the computation of the corresponding eigenmodes in a non-spherical potential is more involved.

\subsection{Target Density Profile}
\label{subsec:NFW+core_density}

For the purpose of illustrating the procedure in this section, we adopt a target density profile corresponding to an NFW halo with a soliton-like core, consistent with the halo density profiles found in cosmological simulations of structure formation \cite{Schive:2014dra,Schive:2014hza,Mocz:2017wlg,Veltmaat:2018dfz}. The outer profile is defined using the classical NFW prescription \citep{Navarro:1996gj}:
\begin{equation} \label{eq:NFW}
    \rho_\text{NFW}(r) = \frac{\rho_0}{(r/r_s)(1 + r/r_s)^2} \ ,
\end{equation}
with a scale radius of $r_s=10$ kpc and a scale density of $\rho_0=1.1\times10^{6}$ M$_\odot$ / kpc$^3$, corresponding to a virial mass $M_\text{vir}\approx10^{10}$ M$_\odot$, enclosed within $r_\text{vir}=56$ kpc (the radius within which the average density of the halo is 347 times the background matter density).\footnote{All the dynamical wave simulations in this work are performed in isolated, periodic boxes, for which the virial radius is an arbitrary cutoff for calculating the halo mass (the mass of an NFW profile out to large radii is logarithmically divergent). However, we quote the virial mass in order to relate to other works (e.g., \citep{Schive:2014hza}) in which cosmological simulations are utilized to investigate the behavior of FDM.}
We truncate the density profile around the virial radius by multiplying Eq. \ref{eq:NFW} by an exponential factor: $\exp(-r^2 / 2r_\text{vir}^2)$, in order to keep the total halo mass finite. This also ensures that our wave function $\Psi$ does not extend beyond the simulation box for the dynamical simulations carried out in \S\ref{sec:sims}. 

We replace the inner density cusp of the NFW profile with an FDM-like core following \cite{Schive:2014hza}. The core density profile is described by an approximation of the soliton solution to the SP equation \citep{Schive:2014dra}:
\begin{equation} \label{eq:schive_core_density}
    \rho_c = \frac{0.019 (m_a / 10^{-22} \text{eV})^{-2} (r_c / \text{kpc})^{-4}}{[1 + 0.091(r / r_c)^2]^8} \ ,
\end{equation}
where we use the core radius given by the scaling relation from \cite{Schive:2014hza} (with $z=0$):
\begin{equation} \label{eq:schive_core_radius}
    r_c = 1.6 \bigg(\frac{m_a}{10^{-22} \text{eV}}\bigg)^{-1} \bigg(\frac{M_\text{vir}}{10^9 \text{M}_\odot}\bigg)^{-\alpha} \text{kpc} \ ,
\end{equation}
with $\alpha=1/3$, yielding a core radius of 0.9 kpc for the density profile discussed in this section.

\begin{figure}[ht!]
\includegraphics[width=\columnwidth]{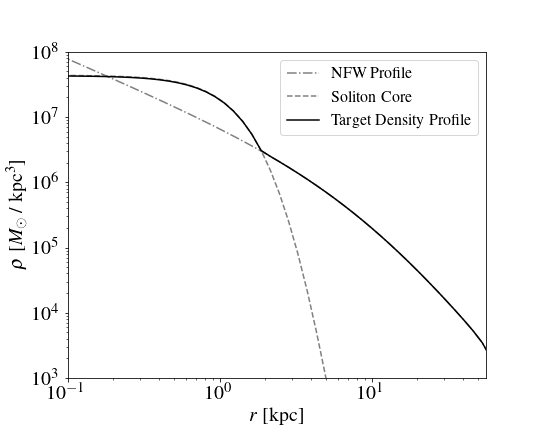} 
\caption{The target density profile for this section, consisting of an NFW halo with a superposed soliton core following the \citet{Schive:2014hza} core-halo relation.
}
\label{fig:2_1}
\end{figure}

The target density profile is shown in Figure \ref{fig:2_1}. The transition between the inner core density profile and the outer NFW profile occurs at approximately $2r_c$.  In \S\ref{sec:profiles} and \S\ref{sec:core-halo} we apply this method to fit a variety of other spherically symmetric density profiles, including a similar NFW profile without a superposed soliton core.

\subsection{Calculation of Eigenmodes}
\label{subsec:library}

Given the spherical symmetry of our target halos, a convenient basis is to factorize each eigenmode into its radial and angular components: $\psi_j = \psi_{n \ell m}(r,\theta,\phi) = R_{n\ell}(r)Y_\ell^m(\theta,\phi)$, where $n$, $\ell$, and $m$ are the radial,\footnote{Many sources use $n$ to instead denote the \textit{principal} quantum number. The radial quantum number
$n$ here corresponds to the number of nodes in $R_{n\ell}$, and is related to the principal number through $n = n_\text{principal} - \ell - 1$.} angular, and magnetic quantum numbers, respectively (i.e., the subscript $j$ is a proxy for $n, \ell, m$), $Y_\ell^m$ are the spherical harmonics, and the radial functions $R_{n\ell}$ are obtained by solving:
\begin{equation} \label{eq:schrod-rad}
    -\frac{\hbar^2}{2m_a}\frac{d^2u}{dr^2} + \bigg[\frac{\hbar^2}{2m_a}\frac{\ell(\ell+1)}{r^2} + m_aV(r) \bigg]u = Eu \ ,
\end{equation}
with $u_{n\ell}(r)\equiv rR_{n\ell} (r)$. The energy eigenvalue $E$ depends on
$n$ and $\ell$ but not $m$, by virtue of spherical symmetry.

Using the density profile described in \S\ref{subsec:NFW+core_density} as the input density profile, we compute the corresponding gravitational potential $V$ and then solve Eq. \ref{eq:schrod-rad} numerically for each value of $\ell$.\footnote{\label{rgrid}The eigenvalue/eigenmode problem is solved numerically on a finite grid from close to the origin out to $2r_\text{vir}$. We have verified that the obtained eigenvalues and eigenmodes are sufficiently accurate, and that extending the grid out to larger radii does not change the results significantly. All eigenmodes are obtained via the finite difference method, using \texttt{SciPy}'s linear algebra library, with the exception of the $\ell=0$ modes, for which we utilize the dedicated boundary value problem solver from the \texttt{SciPy} integration and ODE library, to ensure the accuracy of the $\ell=0$ solutions at small $r$.
}
This yields a series of eigenvalues $E_{n\ell}$ and corresponding radial functions $R_{n\ell}$, that each have $2\ell + 1$ degenerate states once we include the corresponding spherical harmonic term $Y_\ell^m$. The eigenmodes are each normalized such that:
\begin{equation} \label{eq:norm}
    \int r^2 \sin{\theta}\ |\psi_{n \ell m}|^2\ dr\ d\theta\ d\phi = 1 \ .
\end{equation}

\begin{figure}[ht!]
\includegraphics[width=\columnwidth]{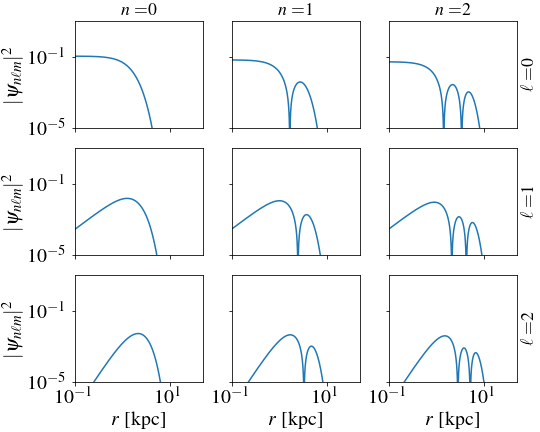} 
\caption{The radial profiles of the first few eigenmodes, organized by their radial ($n$) and angular ($\ell$) quantum numbers. The ground state is shown in the panel on the upper left.
}
\label{fig:2_2}
\end{figure}

Figure \ref{fig:2_2} shows the first few eigenmodes, labeled by the values of $n$ and $\ell$ (noting that the degenerate states with a given value of $n$ and $\ell$ but different values of $m$ do not differ in their radial profiles).

We include all eigenmodes from  the ground state up to a maximum energy, chosen to be the energy corresponding to a particle on a circular orbit at $r_\text{vir}$. Note that the eigenmodes can extend beyond $r_\text{vir}$ (see footnote \ref{rgrid}) -- especially the higher energy ones, but they are also assigned very small amplitudes, by virtue of our construction method where the target density profile is exponentially suppressed beyond $r_\text{vir}$. 

\subsection{Determination of Amplitudes and Application of the Schwarzschild Method to Wave DM Halos} \label{subsec:schwarz}

Self-consistency requires that we choose the amplitude coefficients $a_{n\ell m}$ that correspond to each eigenmode $\psi_{n\ell m}$ such that the constructed wave function reproduces the correct (time-averaged) density profile. In other words, the reconstructed density must source the potential $V$ that was used to construct the eigenmodes in the first place. In what follows, we describe a procedure inspired by Schwarzschild's method, for which the inputs are the desired density profile and a few assumptions about the distribution of amplitudes.

\citet{Schwarzschild1979} devised the following four-step approach to produce stable and self-consistent galactic halos (triaxial in general, and composed of stars or other classical particles):
\begin{enumerate}
    \item Choose an initial density distribution ($\rho_{\text{in}}$).
    \item Compute the gravitational potential corresponding to the density distribution from step 1.
    \item Create a large library of orbits that exist in the potential from step 2, and translate each orbit into a density distribution based on the average amount of time a star on that orbit spends at each location.
    \item Reproduce the original density distribution through a superposition of the orbits calculated in step 3 by assigning each orbit an occupation number (the number of stars on that orbit).
\end{enumerate}
For our wave halo construction, steps 1 and 2 are the same, while step 3 was covered in the preceding subsection: the computation of orbits is replaced by the computation of eigenmodes.

The wave analog of step 4 is to choose a superposition of eigenmodes to reproduce the desired density distribution, i.e., choose the $|a_j|^2$ (where $j$ is a shorthand for $n, \ell, m$) in Eq. \ref{eq:wave-func-diag} such that the time-averaged density profile matches the input. Each $a_j$ comes with a phase, which has no impact on the time-averaged profile but does affect the instantaneous one (Eq. \ref{eq:number-dens}). We assign the phase randomly for each eigenmode, much as in Schwarzschild's construction, where the orbital phase for each particle is chosen randomly.

In Schwarzschild's classic paper, step 4 was formulated as a linear programming problem -- one searches for the superposition of orbits that yields the desired density profile, while minimizing a cost function of one's choosing \citep{Schwarzschild1979, Richstone1984}.
Linear programming was a useful technique because of its low computational cost. With modern-day computational resources, the problem of searching for the right superposition can be thought of as a parameter-fitting problem, where one minimizes some effective $\chi^2$ or maximizes some effective likelihood, just like problems in data analysis.

In the classic Schwarzschild construction, there are typically many possible ways to achieve a given density profile. For instance, one could choose to build a halo out of different combinations of radial and non-radial orbits. The role of the cost function is to break this degeneracy \cite{Richstone1984}):
one could choose to minimize the difference between the tangential velocity dispersion and the radial one (in which case, the constructed halo will have effectively an isotropic velocity dispersion), or one could minimize the tangential velocity dispersion (in which case, the constructed halo will consist of radial orbits).

We have similar freedom in our wave construction. As a first step,
we demand that the amplitude $a_{n\ell m}$ be independent of $m$, as is appropriate
for a spherically symmetric halo: the $|\psi_j|^2$ in Eq. \ref{eq:wave-func-diag} contains
$|Y_\ell^m (\theta,\phi)|^2$ whose sum over $m$ is independent of $\theta$ and $\phi$:
\begin{equation} \label{eq:sph_harm_norm}
    \sum_m |Y_\ell^m (\theta,\phi)|^2 = \frac{2\ell+1}{4\pi} \ .
\end{equation}

We proceed by optimizing $|a_{n\ell m}|^2$ to minimize the following quantity:
\begin{equation} \label{eq:func_to_min}
    D(\rho_\text{in}, \rho_\text{out}) = \frac{1}{r_\text{fit}}\int_0^{r_\text{fit}} dr \bigg(\frac{\rho_\text{out} - \rho_\text{in}}{\rho_\text{in}}\bigg)^2 \ ,
\end{equation}
with $\rho_\text{in}$ representing the target density and $\rho_\text{out}$ representing the average density profile according to
Eq. (\ref{eq:wave-func-diag}), or:
\begin{eqnarray} \label{eq:optimize-amps}
    \nonumber \rho_{\text{out}}(r) & = & m_a \sum_{n\ell m} |a_{n\ell m}R_{n\ell}(r)Y_\ell^m(\theta, \phi)|^2 \\
    & = & \frac{m_a}{4\pi}\sum_{n\ell} (2\ell+1) |a_{n\ell m}|^2 |R_{n\ell}(r)|^2 \ ,
\end{eqnarray}
where we have used Eq. \ref{eq:sph_harm_norm} and the fact that the value of $a_{n\ell m}$ is independent of $m$. We choose $r_\text{fit}$, the maximum radius out to which we attempt to fit the target profile, to be $1.2r_\text{vir}$.

There is some arbitrariness in the choice of $D$
(which, in a way, plays the role of the cost function): for instance, one could choose to give more weight to deviations from the target density profile at smaller or larger radii. 
The overall goal is to produce an output
density profile that is as close to the input as possible, but there is no guarantee the two would be equal. For instance, the output density profile can never match an input profile that is cuspy at small radii.
This is because only $\ell=0$ modes can contribute to the density at $r=0$, and all $\ell=0$ eigenmodes are characterized by a flat (or cored) density profile around $r=0$ (see Figure \ref{fig:2_2}).
Thus, no superposition of modes can possibly create a density cusp at small radii. In this section, we replace the cuspy inner region of the NFW profile with a soliton-like core to circumvent this issue, but we demonstrate the outcome of this procedure on cuspy input profiles in \S\ref{sec:profiles}.

To improve our solution, we repeat the four-step process described above a few times, updating the target density with the previous iteration's output density each time. This is particularly important for cases with cuspy target profiles, where one iteration alone may not lead to a self-consistent solution; the first iteration may lead to an output density profile that deviates considerably from the original input density, meaning the eigenmodes that were used to construct the output in the first iteration may no longer be supported by the output density. Having noted that, it only takes a few such iterations for the method to converge to a self-consistent construction; in all the applications described in this work we find that the solution never requires more than five iterations to converge, and in most cases one or two iterations suffice.

\begin{figure}[ht!]
\includegraphics[width=0.45\textwidth]{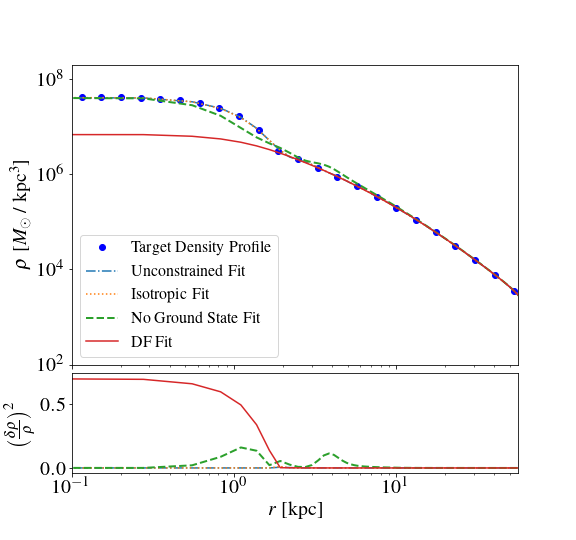} 
\caption{Comparison of the target density profile (blue dots) to the output density profiles from each of the methods described: (1) unconstrained Schwarzschild algorithm (light blue dot-dashed line, almost identical to the target profile), (2) Schwarzschild algorithm with isotropic constraint (orange dotted line, also almost identical to the target profile),  (3) Schwarzschild algorithm with no ground state (green dashed line), and (4) using directly the particle distribution function $f$ of the target density, i.e., $|a_{n\ell m}|^2 = (2\pi\hbar)^3 f /m_a^4 $ (red solid line, labeled DF Fit).
The bottom panel shows the fractional difference (squared) from the target: (1) and (2) match the target profile at all radii; (3) and (4) match the target at large but not small radii.
}
\label{fig:2_3}
\end{figure}

A comparison of the NFW+core input density profile and the self-consistent output of this procedure is shown in Figure \ref{fig:2_3}. We label this the unconstrained fit, because no additional constraints are placed on the amplitudes of modes with different values of $n$ or $\ell$. 
In this case -- and due to the target density profile having a core that is meant to match the ground state -- the method produces an output density profile that matches the target very well. 

Additional constraints can be implemented with ease. Figure \ref{fig:2_3} includes two additional self-consistent outputs obtained by adding various constraints to our wave implementation of the Schwarzschild method:
\begin{enumerate}
    \item An isotropic fit, in which the eigenmodes are binned based on their energy eigenvalues $E_{n\ell}$, and we require the amplitude $a_{n\ell m}$ to be identical for all eigenmodes in the same energy bin. This is the analog of demanding isotropic velocity dispersion for particle orbits (see examples in \cite{Richstone1984}). In the parlance of the particle distribution function $f$, it is the wave analog of imposing $f$ as a function of energy alone. This fit appears to match the target potential to a very high degree of accuracy at all $r$, just like the unconstrained fit.
    \item A fit that does not include the ground state, in which the ground state amplitude is forced to be zero. This is an interesting toy example for studying ground state condensation. This fit appears to do well at large radii, but (unsurprisingly) fails to match the precise shape of the density profile at radii where the solitonic core dominates the target potential.
\end{enumerate}

The output density for each of these options is shown in Figure \ref{fig:2_3}. In some cases, additional constraints can cause the best-fit self-consistent output configuration to deviate from the target density. For example, while the no-ground-state fit succeeds in matching the outer density profile, it struggles to fit the exact shape of the inner soliton because no superposition of excited states is able to exactly reproduce the density profile of the missing ground state.

It is natural to guess that $|a_{n\ell m}|^2$ should be somehow proportional to the particle distribution function $f$. Indeed, this was the proposal by Widrow and Kaiser \cite{Widrow:1993qq}, and was utilized in \citep{Dalal:2020mjw}. This can be made concrete in the simple case where $|a_{n\ell m}|^2$ and $f$ are functions of energy alone. The well-known Eddington formula (Eq. \ref{eddington})  \citep{Eddington1916, BT} can be used to compute $f(e)$ for a given 
spherically symmetric density (and therefore potential) profile, where we use lowercase $e$ to refer to the energy per unit mass $E/m_a$.
As we show in Appendix \ref{sec:app}, in the high energy (i.e., WKB) limit, 
$|a_{n\ell m}|^2 \sim (2\pi\hbar)^3 f /m_a^4$ (Eq. \ref{feai}).\footnote{There are two main results in Appendix \ref{sec:app}. Eq. \ref{aiNe} gives the general relation between $|a_j|^2$ and $f(e)$ without using the WKB approximation (but it does assume a continuum limit such that the number of eigenstates per energy is well-defined). Eq. (\ref{feai}) provides the WKB limit of this relation.}
Figure \ref{fig:2_3} also shows the output density profile for this particular choice of the superposition amplitudes (labeled DF fit), to compare it to the outputs of the Schwarzschild method discussed above.
One can see that the DF fit matches the target profile well at large radii, but fails at small radii (see also \cite{Dalal:2020mjw}). 
This is because the inner region is dominated by eigenmodes at low energies, where
the WKB approximation breaks down. Contrast this with the isotropic fit, which by construction
also employs amplitude coefficients that depend purely on energy, but are allowed to freely 
adjust to fit the target density profile. The isotropic fit does a good job at all radii.

\begin{figure}[ht!]
\includegraphics[width=0.45\textwidth]{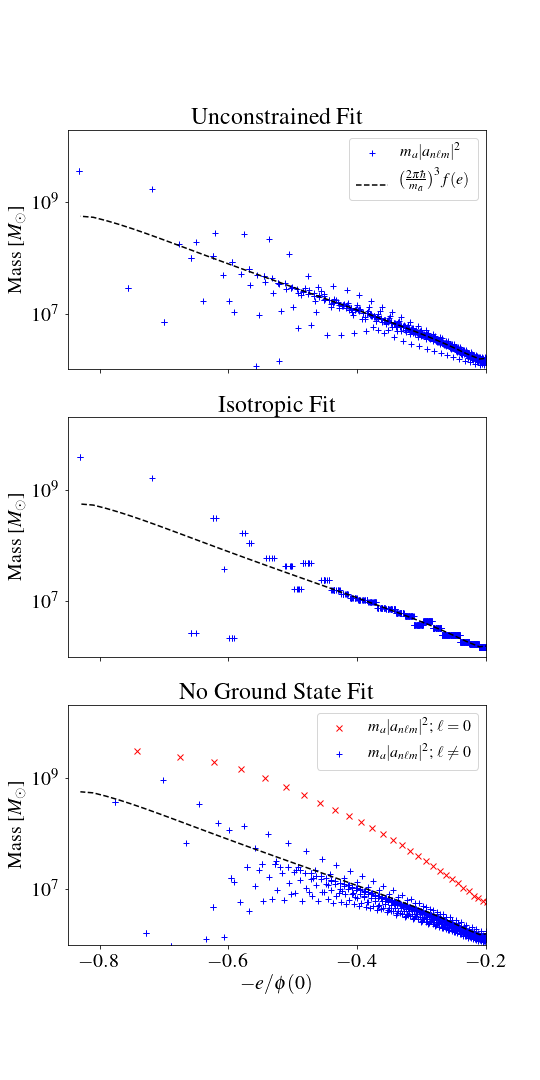} 
\caption{Comparison of the output amplitudes $a_{nlm}$ (for the unconstrained/isotropic/no-ground-state fit in the top/middle/bottom panel) and the distribution function $f$ obtained from the Eddington inversion formula. The blue $+$'s represent $m_a|a_{n\ell m}|^2$, or the total mass found in each eigenstate. The black-dashed line is proportional to the distribution function: $(2\pi\hbar)^3 f(e) /m_a^3$ (the proportionality constant is derived in Eq. \ref{feai} in \S\ref{sec:app}). In the bottom panel (the fit with no ground state), we differentiate between the $\ell=0$ modes and the $\ell>0$ modes by plotting the former with red $\times$'s, in order to show how, absent the ground state, the excited $\ell=0$ modes are all elevated in this fit in order to match the central density of the target density profile (see discussion in the text).
}
\label{fig:2_4}
\end{figure}

To reveal the inner workings of the constructed halos, we show in
Figure \ref{fig:2_4} the squared eigenmode amplitudes $|a_{n\ell m}|^2$, sorted by energy,
and compare them against $f(e)$. We can see that in all cases, most of the 
amplitudes (squared) do roughly have the same energy dependence as $f(e)$, but there
are also eigenmode amplitudes with significant deviations.
The isotropic fit provides the most fair comparison, since the amplitudes
are by construction dependent on energy alone, just as in the distribution function. 
In that case, we still see significant deviations at low energies. 
The unconstrained fit has a larger scatter, reflecting the fact that the amplitudes depend not only on energy, but also on angular momentum.

\begin{figure*}[ht!]
\includegraphics[width=\textwidth]{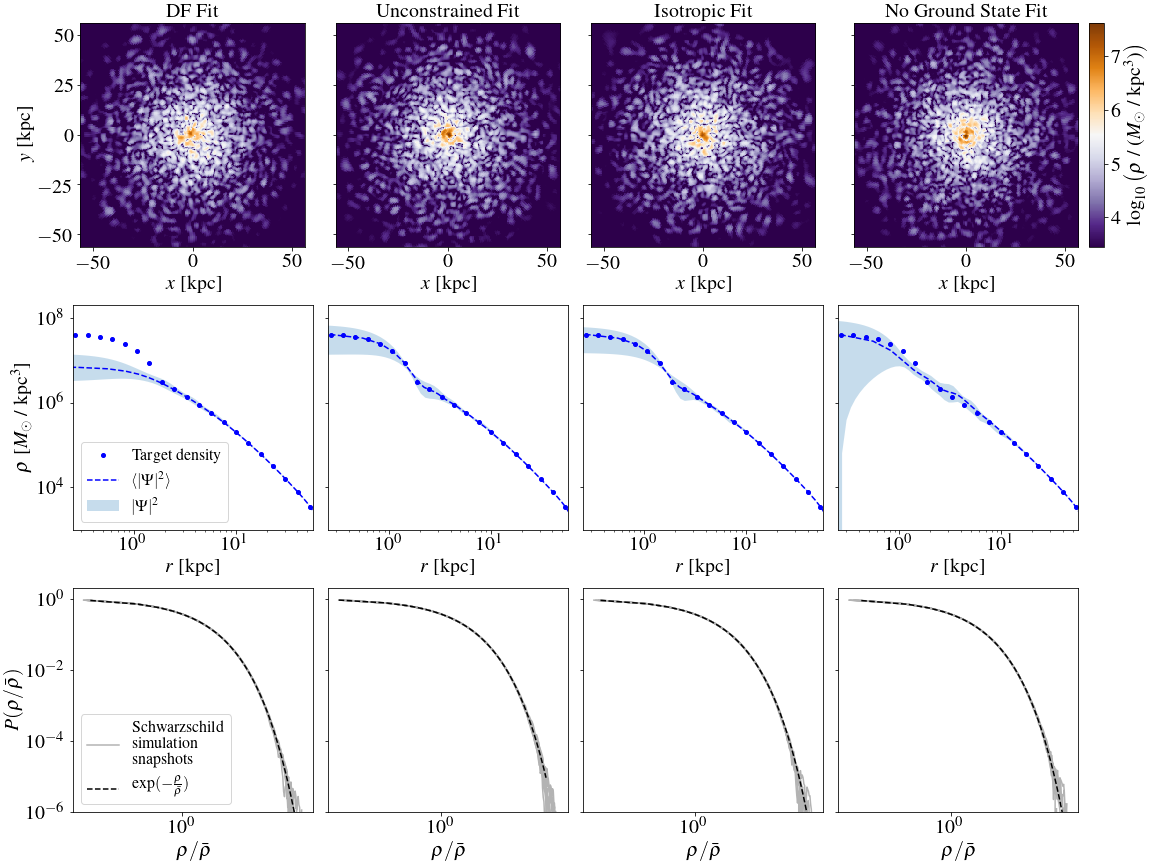} 
\caption{Further details on each of the constructed halos presented in Figure \ref{fig:2_3}. Top row: Density slices along the XY plane at $z=0$ at one particular moment in time. Middle row: 
the blue dots mark the target density profile, the blue dashed line shows the output average density profile from the wave Schwarzschild method, and the blue shaded region depicts the rms density fluctuations from the analytic evolution of the eigenmodes' phases.
Bottom row: the probability distribution of the density fluctuation $\rho/\bar\rho$,
where $\bar\rho$ is the local average density; light gray lines represent the probability distribution measured from 10 different snapshots (one line for each snapshot), and
the black dashed line is the analytic prediction.
}
\label{fig:2_5}
\end{figure*}

In the case of the fit that does not include the ground state (the bottom panel in Figure \ref{fig:2_4}), the lack of a ground state forces the algorithm to assign larger amplitudes to the excited $n>0$, $\ell=0$ eigenstates, in order to match the central region of the target density profile (eigenstates with $\ell>0$ do not contribute to the density at $r=0$, as shown in Figure \ref{fig:2_2}). These are the eigenmode amplitudes plotted in red $\times$'s in Figure \ref{fig:2_4} that are systematically above the $f(e)$ line.

In Appendix \ref{sec:app}, we include a second demonstration of how the WKB approximation can also be used to relate $a_{n\ell m}$ to the distribution function, this time of a halo constructed entirely out of radial ($\ell = 0$) modes (see Figure \ref{fig:app_1}).

\subsection{Random Phases and Halo Evolution} \label{subsec:evolve}

As a final step in the process, we assign each eigenmode a random phase by multiplying $a_{n\ell m}$ by $e^{i\phi_{n\ell m}}$ where $\phi_{n\ell m}$ is a randomly chosen number between $0$ and $2\pi$ (note that unlike $a_{n \ell m}$, $\phi_{n \ell m}$ is also dependent on $m$, i.e., different $m$ modes with the same $n$ and $\ell$ have independently assigned phases). A three-dimensional FDM halo can then be produced. Its time evolution is simple: propagate the phase of each eigenmode analytically, according to Eq. \ref{eq:wave-func-full}. The halos discussed in this section are created from a superposition of $\sim10^5$ eigenmodes, and the efficient numerical construction of these halos is enabled by fast spherical harmonic transforms using the \texttt{SHTools} library\footnote{https://shtools.github.io/SHTOOLS/index.html} \citep{Wieczorek2018}.

The output of the numerical construction of four halos from Figures \ref{fig:2_3} and \ref{fig:2_4} is shown in the four columns of Figure \ref{fig:2_5}. 
Each of the panels in the first row is a density slice through each numerically constructed halo, exhibiting clearly the expected small-scale density fluctuations caused by the interference term in Eq. \ref{eq:number-dens}.

\begin{figure*}[ht!]
\includegraphics[width=\textwidth]{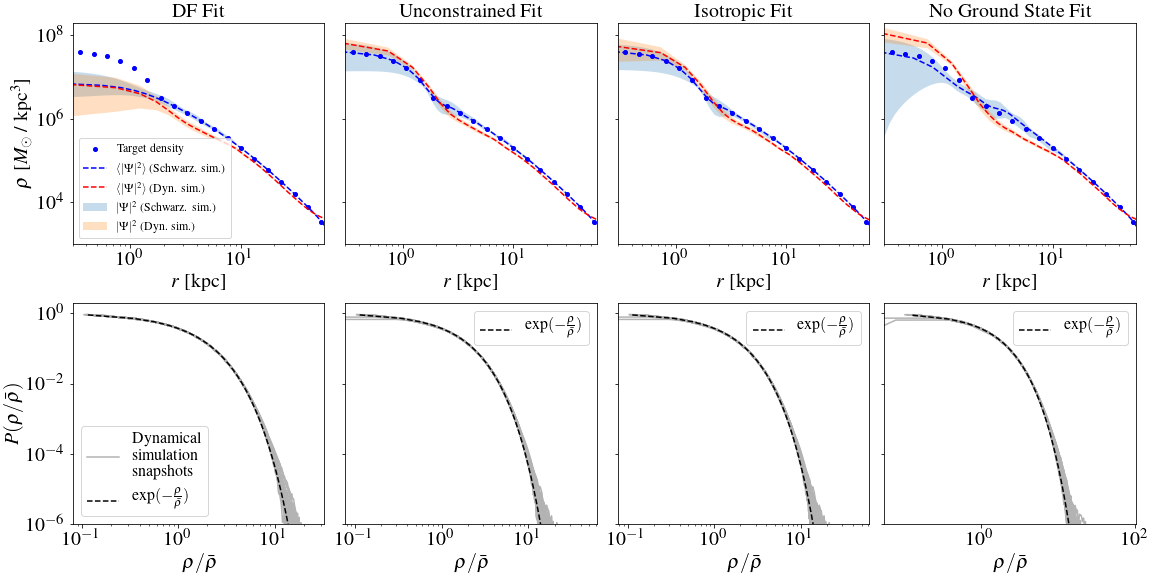} 
\caption{Comparison of the Schwarzschild simulations (i.e., numerical simulations of the Schwarzschild output where the phase of each eigenmode is propagated analytically) to dynamical wave simulations initialized from a snapshot of the output of the Schwarzschild method. Top row: the blue shaded region shows the rms density fluctuations calculated from the Schwarzschild construction, evolved by analytically propagating the eigenmodes' phases (the blue dashed line is its time-average); the orange shaded region shows the rms density fluctuations from the dynamical wave simulations (the red dashed line is its time-average).
The original target density profile is shown in blue dots for reference. 
Bottom row: the one-point probability distribution of the density fluctuation obtained from snapshots of the dynamical wave simulations (comparable to the bottom panels of Figure \ref{fig:2_5}, which are obtained from the Schwarzschild simulations).
}
\label{fig:3_1}
\end{figure*}

The second row of Figure \ref{fig:2_5} illustrates how the density profiles of these numerically constructed halos fluctuate over time (we emphasize the time evolution depicted here comes entirely from the analytic propagation of each eigenmode's phase; dynamical wave simulations would be required if one is interested in the exact time evolution, as discussed in the next section).
The density fluctuations at large radii appear to be less pronounced compared to those at small radii, but that is mostly an artifact of spherical averaging -- the density profile at a larger radius is averaged over shells of a larger volume.

Several differences stand out among the four constructed halos shown in this figure:
\begin{itemize}
    \item In the first three fits shown in Figure \ref{fig:2_5}, the central region is characterized by density fluctuations caused by interference between the ground state and the most prominent excited states, similar to those discussed in several recent works \cite{Veltmaat:2018dfz,Schive:2019rrw,Li:2020ryg,DuttaChowdhury2021}. These density fluctuations are especially prominent in the final fit (the fit with no ground state); in this, the most prominent excited states all have comparable amplitudes (as shown in the bottom panel of Figure \ref{fig:2_4}), leading to large fluctuations.
    \item The unconstrained fit and the isotropic fit both do a better job matching the target central core density compared to the DF fit, for reasons discussed earlier. Moreover, they appear to be nearly identical to each other. This is particularly important from the perspective of computational efficiency; while the unconstrained fit optimizes the amplitudes for all $n$ and $\ell$ eigenmodes ($>1000$ degrees of freedom), the isotropic fit achieves very similar results with only $30$ parameters corresponding to the $30$ bins in energy space into which the eigenmodes are sorted.
    \item The fit without the ground state eigenmode overpopulates some excited states in order to match the desired central density of the halo (see Figure \ref{fig:2_4}). However, overpopulating these excited states still does not fit the core precisely, and also leads to larger density fluctuations outside the core region. It turns out that it is impossible to fit both the inner density core and the outer regions simultaneously without the ground state.
\end{itemize}

Finally, the third row of Figure \ref{fig:2_5} depicts the one-point probability distribution of density fluctuation of each of the halos. Here, $\bar\rho$ refers to the local average density and a histogram of $\rho/\bar\rho$ provides a measure of the probability distribution of the density fluctuation. The light gray lines represent the distribution from 10 different snapshots (one for each snapshot). The black dashed line represents the analytical prediction:
\begin{equation} \label{eq:1-point}
    P(\rho/\bar\rho) = e^{-\rho / \bar{\rho}} \ ,
\end{equation}
i.e., $P(\rho/\bar\rho) \, d(\rho/\bar\rho)$ is the probability that $\rho/\bar\rho$ falls within
the indicated range. 
This probability is sometimes referred to as the Rayleigh distribution \cite{Centers:2019dyn}, and can be derived from the assumption of random phases for the eigenmodes
\cite{Hui:2020hbq}. 
The measured probability distribution agrees very well with the analytic prediction.

As explained earlier, the Schwarzschild wave construction involves an approximation: the eigenmodes are computed using the time-averaged gravitational potential; gravitational potential fluctuations due to the interference-induced substructures are ignored. The assumption is that the substructures have negligible effects on the global structure of the halos. In the next section, we test this assumption by performing dynamical simulations of wave halos 
(i.e., solving the Schr\"odinger-Poisson system self-consistently) that are initialized from a snapshot of our constructed halos, and by comparing these dynamical simulations to the evolution of the halos in our Schwarzschild simulations.

\section{Dynamical Simulations and Halo Decomposition} \label{sec:sims}

We now turn to testing the stability of the halos constructed using the modified Schwarzschild method, by comparing the evolution of the Schwarzschild simulations described in the previous section to dynamical wave simulations. To do so, we take a snapshot of each of the halos constructed in the previous section, and evolve it by numerically solving the Schr\"odinger-Poisson system, using the \texttt{SPoS} code described in \cite{Li:2018kyk}.

Each dynamical simulation is performed in a $256^3$ box whose sides measure $2r_\text{vir} = 113$ kpc, with a spatial resolution of $\sim0.4$ kpc, and periodic boundary conditions. The chosen box size ensures that the density at the edges is 3-4 orders of magnitude smaller than the central density. As a result, the periodic boundary conditions do not have a significant impact on the evolution of the halo. Note that none of these are cosmological simulations, and the background density is not taken into account.

In order to initialize each dynamical simulation, we project the output wave functions from the previous section onto the dynamical simulation grid as initial conditions. Each halo is then evolved for 16 Gyr, or approximately eight free-fall times, defined as:
\begin{equation} \label{eq:t_ff}
    T_{\text{ff}} = \sqrt{\frac{\pi^2r_{\text{vir}}^3}{8MG}} \ .
\end{equation}

The top row of Figure \ref{fig:3_1} compares the density profiles of the constructed halos against the profiles measured from the dynamical simulations.
The blue shaded regions depict the Schwarzschild constructions, evolved 
by analytically propagating the eigenmodes' phases 
(identical to those shown in Figure \ref{fig:2_5}); the blue dashed lines show the corresponding time averages.
The orange shaded regions show the range of density profiles measured from
snapshots of the dynamical wave simulations; the red dashed lines show the corresponding averages.
All dynamical simulation snapshots are taken from after the halo has evolved for 4 Gyr, in order to give the halos ample time to relax into a steady state. Significant deviations of the dynamical simulations from the Schwarzschild 
constructions would thus indicate that the Schwarzschild method has failed to produce a globally stable halo.

At first glance, the halo initialized directly from the distribution function appears to match the target profile well, though a more careful inspection reveals that the dynamical simulation relaxes to a somewhat different configuration with a core that is slightly more prominent than the rest of the halo.

In contrast, the halos initialized from both the unconstrained fit and the isotropic fit remain stable throughout the duration of the dynamical halo evolution, maintaining an average density profile that is similar to the output time-averaged density from the Schwarzschild method. In both halos, the core grows by a small fraction and the density fluctuations at the core become slightly less pronounced, but overall the deviations of the dynamical simulations from the Schwarzschild construction are small. This suggests, at least for our chosen halo mass and target density profile, that the Schwarzschild construction method is able to produce stable and self-consistent halos. Moreover, the similarity between the results in these two columns further supports the assertion that the isotropic method can be used to obtain similar results as the unconstrained method, at a fraction of the computational cost.

The halo initialized without a ground state does not maintain its original configuration -- the time-averaged density profile in the dynamical simulations clearly suggest that a prominent soliton has grown for this halo where there wasn't one originally, so much so that it surpasses the central density of the original target density profile by a considerable amount.

The second row of Figure \ref{fig:3_1} shows the probability distribution of density fluctuations measured from the dynamical simulations. In all cases, the probability distribution matches that shown in Figure \ref{fig:2_5} for the Schwarzschild simulations. This suggests that the wave interference substructures seen in the dynamical simulations 
are similar to those in the Schwarzschild simulations, at least as far as the one-point probability distribution is concerned -- we will have more to say about this below.

The dynamical wave simulations allow us to test the Schwarzschild constructions in more detail. For instance, an important assumption of the construction method is that the amplitude for each eigenmode is time-independent, and that the only time-dependence is in the phase.
To test this, we can decompose the dynamically evolved wave function by leveraging the orthonormal nature of the eigenmodes:
\begin{equation} \label{eq:decompose}
    a_{nlm} = \int d^3 x\ \Psi(\bm{x},t)\psi_{nlm}^*(\bm{x}) \ ,
\end{equation}
where $\psi_{nlm}$ refers to the eigenmodes used to obtain the final output potential in \S\ref{sec:method} (i.e., the blue dashed lines in Figure \ref{fig:2_5}).\footnote{In theory, it would be more accurate to decompose the halo at each snapshot of the dynamical simulation by first calculating a new library of eigenmodes based on the exact gravitational potential at that snapshot. However, the density profile doesn't change enough throughout any of the simulations to cause significant inaccuracies due to this issue.
}

\begin{figure}[ht!]
\includegraphics[width=\columnwidth]{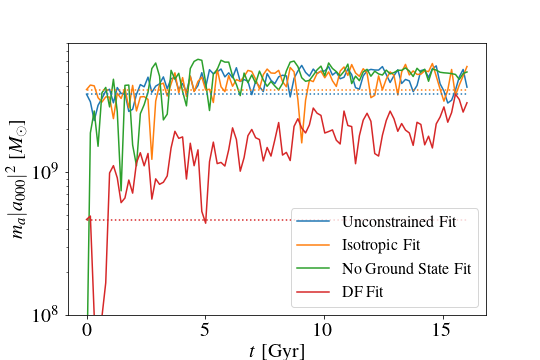} 
\caption{Evolution of the squared amplitude of the ground state mode over eight free-fall times in the four dynamical wave simulations shown in Figure \ref{fig:3_1} (solid lines) compared to their initial amplitudes (shown for reference in dotted lines of the matching color). The ground state amplitudes of the simulations initialized from the unconstrained and the isotropic fits remain relatively unchanged throughout the entire simulation, while the ground state amplitude of the simulation initialized directly from $f(e)$ grows gradually from its original value. Lastly, the Schwarzschild construction in which the ground state is completely depopulated is shown to be unstable in this figure -- the amplitude of the ground state grows rapidly at the very beginning of the dynamical simulation until it reaches a steadier state near the amplitudes of the ground states from the other three halos.
}
\label{fig:3_2}
\end{figure}

\begin{figure*}[ht!]
\includegraphics[width=\textwidth]{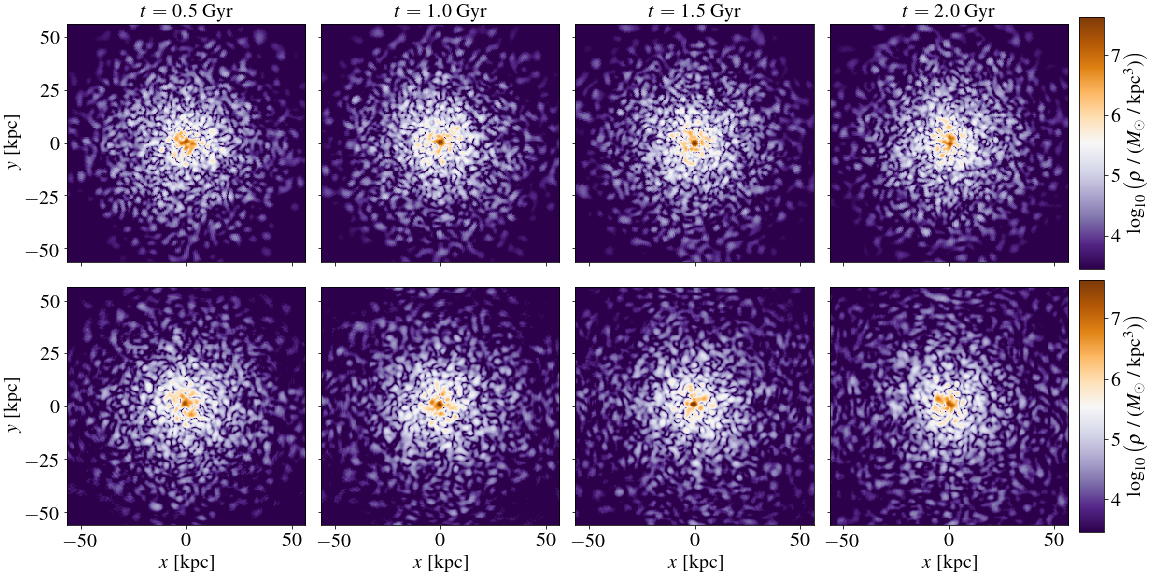} 
\caption{Top row: density slices through four snapshots of the halo constructed using the isotropic implementation of the Schwarzschild method and evolved using Eq. \ref{eq:number-dens} (i.e., analytic propagation of each eigenmode's phase). Bottom row: density slices through four snapshots of the dynamical wave simulation, initialized using the $t=0$ Gyr snapshot of the halo in the top row. Note the similarity of the substructures in the two different kinds of simulations, both in terms of the typical size scale and the time variability. 
}
\label{fig:3_3}
\end{figure*}

Figure \ref{fig:3_2} shows the amplitude of the ground state for each of the four dynamical simulations, evaluated by decomposing each dynamical simulation at various snapshots using Eq. \ref{eq:decompose} for $a_{000}$. The ground state of the halo initialized directly from the distribution function $f(e)$ appears to grow gradually over the course of the dynamical simulation, such that by the end of the simulation the mass in the ground state has nearly doubled (likely at the expense of several of the excited states). For the unconstrained and the isotropic fits, the initial ground state amplitude stays more or less constant throughout the dynamical simulation, never exceeding the original value by more than 50\%, indicating that in both cases the Schwarzschild method has produced not only a globally stable halo, but also one with a stable ground state.

On the other hand, it is clear from Figure \ref{fig:3_2} that the Schwarzschild construction with no ground state is unstable. The ground state is rapidly populated in the dynamical simulation, reaching an amplitude comparable to the other constructions after approximately 100 Myr. This timescale is remarkably close to the crossing time at the core radius for this soliton, which is approximately 80 Myr. The relaxation process in the core that leads to the re-population of the ground state in this last simulation is in line with prior theoretical and numerical work on the topic of self-gravitating Bose-Einstein Condensates (see, e.g., \cite{ss94, Guzman:2006yc, Bernal:2006it, Hui:2016ltb,Levkov:2018kau}).

Finally, the similarities between the Schwarzschild simulations and the dynamical wave simulations are also readily apparent from a qualitative comparison of the two. Figure \ref{fig:3_3} compares four snapshots of each simulation; both rows exhibit similar substructures (both in terms of spatial and time scales), with notable differences only at the edges and particularly the corners of the box.
The two methods are not meant to produce the exact same evolution, of course, but the visual similarities suggest the (much more efficient) Schwarzschild method is a reliable way to statistically study and explore wave substructures.

Given the findings above, we conclude that the Schwarzschild method successfully produces stable and self-consistent wave halos, with the requisite substructures, at least in the case of a \citet{Schive:2014hza} target density profile. 

\section{Application to Other Density Profiles} \label{sec:profiles}

Having focused so far on one type of density profile, and specifically one that is known to remain stable in FDM simulations, we now turn to applying the halo construction method described in \S\ref{sec:method} to a variety of other (spherical symmetric) density profiles. 
It is worth noting that realistic galactic halos exhibit a variety of profiles, due in part to feedback processes, and it is thus useful to be able to construct halos that span a range of profiles.\footnote{One ingredient that will be missing in our construction is baryons or stars. In principle, our approach can be adapted to construct wave dark matter halos of a certain profile, together with baryons following another profile. We leave this for future work.}
We begin this section with two central questions in mind:
\begin{enumerate}
    \item Is it possible to match any arbitrary density profile with an FDM halo using the Schwarzschild halo construction method? If not, what constraints exist?
    \item Do all halos constructed using the wave Schwarzschild method remain as stable in simulations as those simulated in \S\ref{sec:sims}, or does the \citet{Schive:2014dra} NFW+core density profile represent a uniquely stable density profile for FDM halos?
\end{enumerate}

Following the process described in \S\ref{sec:method}, we construct four new halos following four commonly used density profiles in galactic dynamics whose outer density profiles follow different slopes:

\begin{enumerate}
    \item An isothermal (logarithmic) halo, in which the circular velocity is constant and $\rho\propto r^{-2}$ (we choose $v_c=28$ km/s, to match the approximate characteristic circular velocity of the $10^{10}$ M$_\odot$ NFW halos constructed in \S\ref{sec:method}):
    
\begin{equation} \label{eq:Isothermal}
    \rho(r) = \frac{v_c^2}{4\pi Gr^2} \ .
\end{equation}
    
    \item An NFW halo, whose density profile transitions from $\rho\propto r^{-1}$ to $\rho\propto r^{-3}$ at a scale radius of 10 kpc (see Eq. \ref{eq:NFW}). Unlike \S\ref{sec:method} and \S\ref{sec:sims}, here we do not impose a core, meaning the central region of the halo forms a density cusp.
    \item A Hernquist halo, whose density profile transitions from a $\rho\propto r^{-1}$ inner region to $\rho\propto r^{-4}$ at a scale radius of 10 kpc. Like both the isothermal and the NFW profiles, this halo has a central density cusp:
    
\begin{equation} \label{eq:Hernquist}
    \rho(r) = \frac{M_h}{2\pi r_s^3} \frac{1}{(r/r_s)(1+r/r_s)^3} \ .
\end{equation}
    
    \item A Plummer halo, whose density profile transitions from a flat density core to $\rho\propto r^{-5}$ at the scale radius, which we again set at 10 kpc:
\end{enumerate}

\begin{equation} \label{eq:Plummer}
    \rho(r) = \frac{3M_h}{4\pi r_s^3} \bigg(1+\frac{r^2}{r_s^2}\bigg)^{-\frac{5}{2}} \ .
\end{equation}

\begin{figure}[ht!]
\includegraphics[width=\columnwidth]{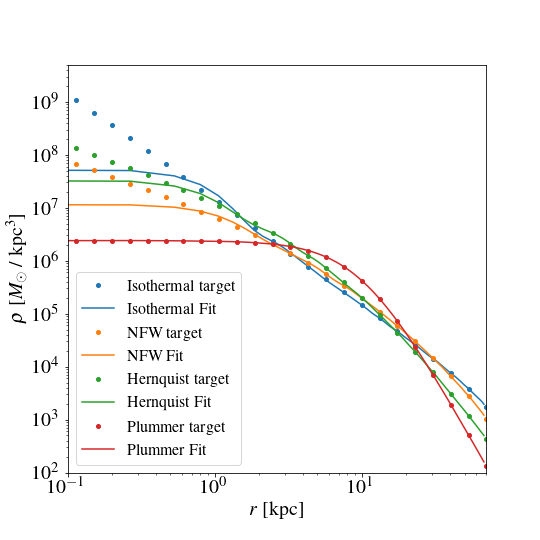} 
\caption{Four target density profiles with $M=10^{10}$ M$_\odot$ (dots), and the output density profile of the Schwarzschild method for each target (solid lines).
}
\label{fig:4_1}
\end{figure}

\begin{figure*}[ht!]
\includegraphics[width=\textwidth]{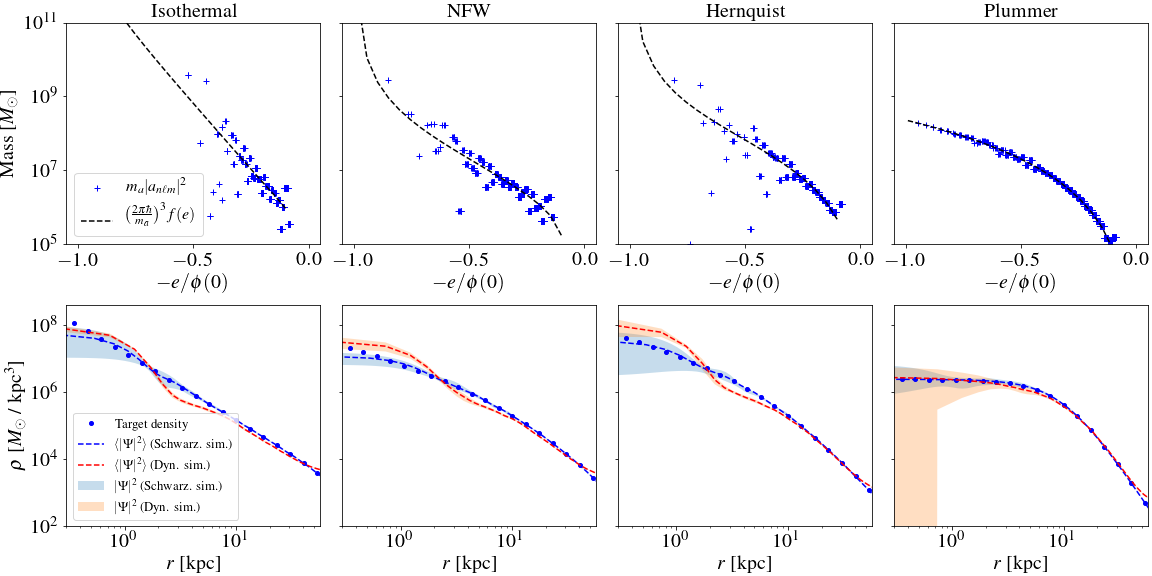} 
\caption{Top row: Output amplitudes for each of the four target profiles shown in Figure \ref{fig:4_1}, plotted against $f(e)$ obtained from the inversion formula. Bottom row: Comparison of the Schwarzschild simulations (i.e., numerical simulations of the Schwarzschild output where the phase of each eigenmode is propagated analytically) to a dynamical wave simulation initialized from a snapshot of the output of the Schwarzschild method, for each of the four halos discussed in this section (see the description of the top row in Figure \ref{fig:3_1} for the meaning of the different lines).
}
\label{fig:4_2}
\end{figure*}

The four target density profiles are shown in Figure \ref{fig:4_1}. Note that both the isothermal and the NFW halos must be truncated at a certain radius to have a finite total mass, which we accomplish in the same manner described in \S\ref{subsec:NFW+core_density}. We retain only the eigenmodes whose energy is lower than the classical circular orbit at $r = 56$ kpc.

As in \S\ref{sec:method}, we find no significant differences between the unconstrained implementation and the isotropic implementation of the Schwarzschild method for any of these halos, so we proceed with only the isotropic results, which are obtained at a significantly lower computational cost. The time-averaged fits obtained from the isotropic implementation of the Schwarzschild method are shown in solid lines in Figure \ref{fig:4_1} (alongside the target density profiles shown in the corresponding dotted lines). 
The wave solutions fit the outer regions of all four profiles to a very high degree of accuracy, suggesting that it may be possible to construct FDM halos with outer density profiles ranging from $\rho\propto r^{-2}$ to $\rho\propto r^{-5}$. Unsurprisingly, the wave solutions fail to fit the central regions of all three cuspy profiles (isothermal, NFW, and Hernquist), while the cored center of the Plummer profile is fit with ease.

Of course, none of these fits is guaranteed to describe a stable halo when the gravitational effect of the time-dependent interference is accounted for in a self-consistent manner. Thus, we turn next to checking whether these numerically constructed halos remain stable in \texttt{SPoS} simulations.

After obtaining the amplitudes $a_{n\ell m}$ for each of the halos and assigning a random phase to each eigenmode, we evaluate the stability of each solution by comparing its evolution using Eq. \ref{eq:wave-func-full} (i.e., analytically propagating the phase of each eigenmode) to the evolution in dynamical simulations (i.e., using the \texttt{SPoS} code described in \S\ref{sec:sims}) initialized from a snapshot of the Schwarzschild-constructed halo.

Figure \ref{fig:4_2} shows a summary of the results for the four halos. The panels of the top row show that, as expected, the amplitudes obtained from the Schwarzschild method largely adhere to the particle distribution functions (calculated from the target density profiles using the Eddington inversion formula), with some deviations appearing at low energies, where the WKB approximation breaks down.

The bottom row compares the Schwarzschild constructions (with phases of eigenmodes evolved analytically) against their dynamical simulation counterparts (this is the same comparison as the one shown in the top row of Figure \ref{fig:3_1} in \S\ref{sec:sims} for the NFW+core profile). In general, the two agree well in the outer regions of all the profiles (out to where the box edges begin affecting the density profile in the dynamical simulations). For the three cuspy profiles, the dynamical simulations appear to converge to a slightly higher central density than that arrived at by the Schwarzschild method. However, as shown in \S\ref{sec:method} this can be easily overcome by forcing a target profile that takes into account something like the core-halo mass relation. The cored center of the Plummer profile remains stable, though even here one may note a \textit{slightly} more pronounced soliton-like core that appears in the dynamical simulation at a smaller radius than the Plummer halo scale radius.

It is worthwhile to pause here in order to appreciate a characteristic of self-consistent FDM halos that can be easily understood through the Schwarzschild construction method. Figure \ref{fig:4_1}  clearly demonstrates that FDM cannot reproduce cuspy halos to arbitrarily small radii. 
At those radii the density profile is instead dominated by the ground state which has a core (in all the panels in the top row of Figure \ref{fig:4_2} the ground state always has the highest amplitude). It raises an interesting question: why not use the excited states to steepen the inner density profile? (By excited states, we mean the $\ell=0$ excited modes; this is because only $\ell=0$ modes contribute appreciably to the inner density). Observe that, for instance from Figure \ref{fig:2_2}, the radial profile of the $n=1$, $\ell=0$ mode prior to its first node, is more compact than the ground state. Why doesn't the Schwarzschild method assign this and other excited $\ell=0$ modes a greater amplitude than the ground state in order to fit the cuspy regions of the target density profiles?

\begin{figure}[ht!]
\includegraphics[width=\columnwidth]{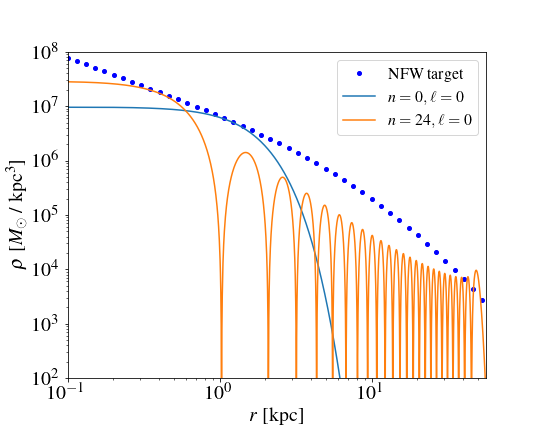} 
\caption{A demonstration of why excited eigenmodes with $\ell=0$ cannot be used to improve the fit to a cuspy target profile without overshooting the outer density profile. The target density profile shown here is the NFW profile used in \S\ref{sec:profiles}, and the two eigenmodes shown are calculated using the corresponding gravitational potential. While the selected excited state is technically capable of fitting the cuspy profile to a smaller radius than the ground state, this cannot be done without contributing to
an outer profile that is shallower than the target density profile.
}
\label{fig:4_3}
\end{figure}

Figure \ref{fig:4_3} illustrates the reason:
the overall shape of these excited modes makes it impossible to
produce a cuspy interior while at the same time fit the outer target profile.
If one were to elevate the amplitude of one of these excited states to the level it would need to fit the central density cusp, one would overshoot in the density in the outer region of the halo. The same can be shown for the other two cuspy profiles discussed in this section; in fact, something similar is also apparent in the fit with no ground state in \S\ref{sec:method}, where the $n=1$, $\ell=0$ mode provides the greatest contribution to the central density but simultaneously leads to a slight overshooting of the target density just outside the core.\footnote{The isotropic constraint imposed in our construction in this section further limits the possibility of assigning the excited $\ell=0$ states a higher amplitude, as that would also necessitate increasing the amplitude of all other states in that given energy bin, which would lead to even larger deviations from the target density profile at large radii. However, Figure \ref{fig:4_3} demonstrates that even without the isotropic constraint, the excited $\ell=0$ modes cannot help with fitting the interior region of a density cusp.}

To conclude this section, we return to the two questions posed at the beginning of the section:
\begin{enumerate}
    \item While a wide variety of spherical profiles can be constructed with a wave-like FDM solution, it is not possible to fit any arbitrary density profile with FDM (of a fixed mass $m_a$). No wave-like solution will successfully fit a density cusp in the regime where $r$ is smaller than the de Broglie wavelength. Furthermore, the central density will be dominated by the ground state: $\ell>0$ modes vanish at $r=0$, and while the excited $\ell=0$ modes do contribute to the central density, their amplitude is constrained by the fact that they also contribute to the outer density profile, as shown in Figure \ref{fig:4_3}.
    
    \item The wave Schwarzschild method is able to fit the outer portion of a wide variety of target density profiles, ranging at least from ($\rho \propto r^{-2}$ to $\rho \propto r^{-5}$), and the constructions are stable as verified by dynamical simulations. While cosmological FDM simulations may lead to predominantly \citet{Schive:2014dra} NFW+core density profiles, it is certainly possible to construct self-consistent and stable FDM halos that follow other density profiles. This is particularly important given that density profiles can be altered by feedback processes, of relevance when comparing theoretical expectations against observations.
\end{enumerate}

\section{Preliminary Investigation of the Core-Halo Relation} \label{sec:core-halo}

As a final demonstration of this method, we turn to a preliminary investigation of the FDM core-halo relation. From \S\ref{sec:method} and \S\ref{sec:sims}, we already know that constructing FDM halos with NFW+core density profiles that follow the \citet{Schive:2014hza} core-halo relation leads to stable, self-consistent solutions. Furthermore, the halos in \S\ref{sec:profiles} demonstrated that `blindly' trying to fit cuspy density profiles can lead to halos whose central regions find a new equilibrium in simulations, even as the outer regions of those halos remain stable at the original target density profile (which does not necessarily have to be NFW-like).

\begin{figure}[ht!]
\includegraphics[width=\columnwidth]{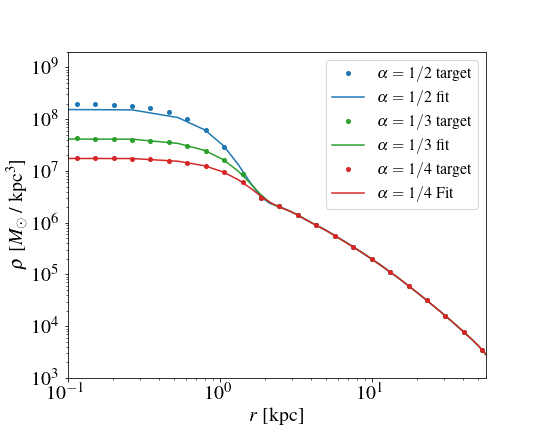} 
\caption{Three target $10^{10}$ M$_\odot$ NFW halos with superposed solitonic cores following three different core-halo mass relations (dotted lines) and the time-averaged density profiles of the Schwarzschild constructions for each of the targets (solid lines).
All three constructions provide very good fits to the target density profiles.
}
\label{fig:5_1}
\end{figure}

\begin{figure*}[ht!]
\includegraphics[width=\textwidth]{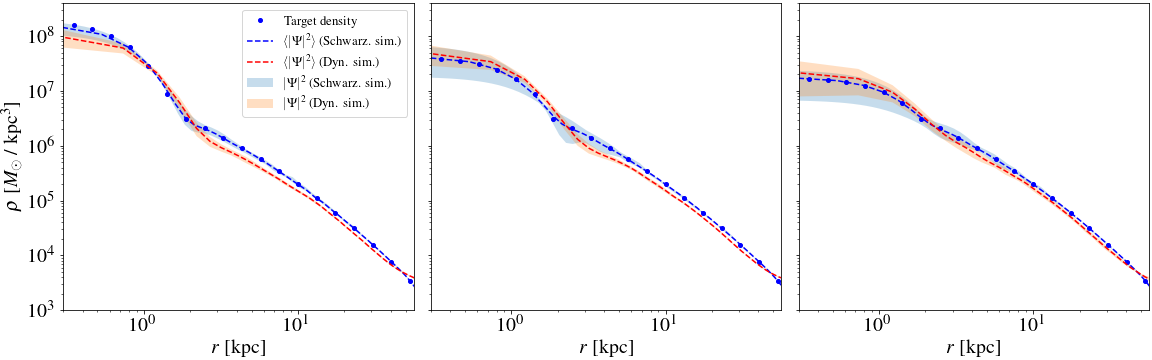} 
\caption{Comparison of the Schwarzschild simulation to the dynamical simulation initialized from a snapshot of the output of the Schwarzschild method, for each of the three halos discussed in this section
(from left to right: $\alpha = 1/2, 1/3, 1/4$). 
See the description of the top row in Figure \ref{fig:3_1} for the meaning of the different lines. 
}
\label{fig:5_2}
\end{figure*}

\begin{figure}[ht!]
\includegraphics[width=\columnwidth]{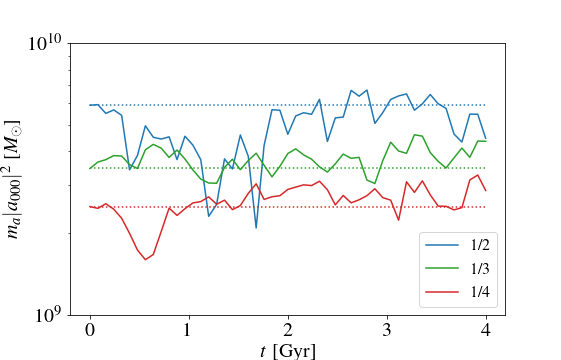} 
\caption{Evolution of the amplitude (squared) of the ground state mode in each of the three dynamical wave simulations described in \S\ref{sec:core-halo}. The ground state amplitudes of all three cases appear to maintain their relative stability with respect to each other. For the $\alpha=1/2$ case, there appears to be an initial period of instability, but eventually the ground state amplitude appears to converge to its original target value after a full free-fall time has elapsed.
}
\label{fig:5_3}
\end{figure}

In this section, we investigate whether other prescriptions for the core-halo relation can also lead to stable solutions. In \S\ref{sec:method} and \S\ref{sec:sims} our target density profile followed the \citet{Schive:2014hza} scaling relation (see Eq. \ref{eq:schive_core_radius}, in which $\alpha = 1/3$). The halos constructed in \S\ref{sec:method} demonstrated that this relation yields self-consistent halos whose cores remain stable over long timescales in simulations. We now compare these halos to two other halos, constructed to fit target density profiles in which $\alpha = 1/2$ and $\alpha = 1/4$ while maintaining the density profile of the soliton solution in Eq. \ref{eq:schive_core_density}. The core radius is inversely proportional to the core mass, so as $\alpha$ grows, the core becomes more massive 
and compact.

It is worth noting that the purpose of this section is not to prove a particular core-halo relation or to place constraints on the value of $\alpha$ -- we leave that investigation to future work. Rather, our goal is simply to see if it is possible to construct stable FDM halos with core-halo relations
that deviate from the one found by \citet{Schive:2014hza}.

As in the previous section, we use the isotropic implementation of the Schwarzschild method to solve for the best-fit amplitudes $a_{n\ell m}$. The solid lines in Figure \ref{fig:5_1} compare the outputs of the Schwarzschild method to the three target profiles (in dotted lines). All three fits converge well to the target profile, with only slight deviations noticeable at small radii in the $\alpha=1/2$ case.

To test the stability of these constructed halos, we again utilize dynamical simulations. As in the previous sections, the dynamical simulations are initialized from snapshots of the three halos constructed using the Schwarzschild method. We run each simulation for two free-fall times (or approximately four Gyr). In Figure \ref{fig:5_2}, we compare the evolution of the Schwarzschild simulations to dynamical wave simulations. 
In all three cases, the dynamical simulations roughly match the Schwarzschild constructions both in terms of the stable time-averaged density profile and in terms of the time-dependent fluctuations, although in the $\alpha=1/2$ case, the dynamically evolved halo deviates a bit more from the constructed one.

Based on Figure \ref{fig:5_2}, there does not seem to be any preference for a specific value of $\alpha$ or $r_c$ for an NFW halo of a given mass. It is important to note that this does not contradict the finding from \citet{Schive:2014hza} that cosmological simulations lead to a core-halo relation with $\alpha=1/3$; it merely suggests that other core-halo relations are also stable. One possible driver of the $\alpha=1/3$ relation may be the specific merger history inherent in dark matter only cosmological simulations such as those described in \citet{Schive:2014hza}. 
Critically, the stability of the cores in all three of the dynamical simulations shown in Figure \ref{fig:5_2} opens the window to consider FDM halos with a variety of core-halo relations, perhaps due to feedback processes or some external perturbations.

To further validate these results, we decompose the halos from the three dynamical simulations shown in Figure \ref{fig:5_2} and evaluate the stability of the ground state amplitude using Eq. \ref{eq:decompose}. The results are shown in Figure \ref{fig:5_3}. All three halos appear to have stable ground states at their respective amplitudes. In the $\alpha=1/2$ case, the beginning of the simulation is characterized by larger amplitude fluctuations, which appear to dampen after approximately one free-fall time, with the ground state amplitude returning to its initial value.

Figures \ref{fig:5_2} and \ref{fig:5_3} suggest that there is no unique ground state amplitude for an FDM halo that follows a given NFW (outer) density profile. 
Rather, it appears that there may be some freedom in the mass contained within the ground state. 

It is worthwhile to note that the results of the experiment described in this section are not directly comparable to those discussed in \S\ref{sec:sims}, because the setup of the experiments in each section is fundamentally different. In \S\ref{sec:sims}, several simulations were initialized from Schwarzschild constructions based on the \textit{same} target density profile. Specifically, initializing a halo to match that cored target density profile, using a superposition of only excited states, led to a halo that was initially out of dynamical equilibrium (as shown in the right-hand column of Figure \ref{fig:3_1} and in Figure \ref{fig:3_2}) -- it is not altogether surprising that the halo relaxed back to its equilibrium state by rapidly repopulating the ground state. On the flip side, in this section we have experimented with three \textit{different} target density profiles, corresponding to three different values of $\alpha$. In each of these cases, the Schwarzschild method appears to find a stable superposition of eigenmodes.

\section{Conclusions} \label{sec:disc}

The main aim of this paper is to lay out an efficient and accurate method for constructing and evolving halos composed of wave dark matter, at a fraction of the computational cost of dynamical wave simulations. A second aim is to clarify the relation between wave superposition amplitudes and the particle distribution function (Appendix \ref{sec:app}).
We have focused on simulations of ultra-light (fuzzy) dark matter for which $m_a=8.1\times10^{-23}$ eV, but the method described in this paper is broadly applicable to wave dark matter at any scale.\footnote{For much larger values of $m_a$, the resulting de Broglie wavelength would be so short that it would be impractical to simulate a whole halo. One could adapt our method to simulate some restricted region of the halo. This is relevant for the study of wave interference substructures in axion detection experiments. See \cite{Hui:2021tkt} for further discussions.}

We adapt the Schwarzschild method for the construction of stable and self-consistent halos -- instead of particle orbits, we seek a suitable superposition of wave eigenmodes that satisfy the time-independent Schr{\"o}dinger equation. The constructed halo can then be evolved by analytically propagating the phases of the eigenmodes, providing a computationally efficient way to simulate the time-dependent wave interference substructures. We verify that this method produces reliable realizations by comparing them to dynamical wave simulations. We find that the constructed halos maintain their mean density profiles throughout the duration of these dynamical simulations (except for somewhat artificial cases, such as the one where we zero out the ground state). 

Along the way, the Schwarzschild construction method allows us to demonstrate several intrinsic properties of halos composed of FDM:

\begin{enumerate}

    \item The outer envelopes of FDM halos can take on a variety of density profiles. In those regions, the dominant eigenmodes have high energies, for which the WKB approximation holds. The amplitudes of those eigenmodes are expected to go as the square root of the particle distribution function (Eq. \ref{feai}), and indeed the Schwarzschild construction confirms that.

    \item The innermost regions of FDM halos must be cored. The central core is dominated by the ground state and the first few $\ell=0$ excited states. It is impossible to create a cusp by some judicious superposition of the eigenmodes.

    \item The central region of the halo is dominated by the low energy eigenmodes, for which the WKB approximation breaks down. Thus, their amplitudes deviate significantly from the classical expectation based on the particle distribution function (compare Eqs. \ref{aiNe} and \ref{feai}). In particular, the ground state amplitude tends to be larger than the classical expectation. We have also verified that a halo in which the ground state is artificially zeroed out, when dynamically evolved self-consistently, would grow a ground state.

    \item Multiple stable amplitudes of the ground state appear to be possible for a halo of a given mass and outer density profile. This suggests that the core-halo mass relation extracted from cosmological dark matter only simulations is not forced upon us by dynamical consistency; rather, it is likely the result of the particular merger history of the model. Feedback processes or external perturbations thus have the potential to alter the relation between the core and the host halo.
    
\end{enumerate}

Several of the topics discussed above warrant further investigation, including the core-halo relation and the timescale for FDM core buildup \cite{Schive:2014dra,Schive:2014hza,Mocz:2017wlg,Veltmaat:2018dfz}, and higher-order statistics for characterizing the wave interference substructures in FDM halos (such as the 1-point statistics shown in Figures \ref{fig:2_5} and \ref{fig:3_1}).
It is worth stressing that the Schwarzschild method is designed for constructing stable, virialized halos. For truly dynamical situations -- where the amplitudes of eigenmodes are not constant, and the gravitational potential evolves significantly -- a
Poisson-Schr\"odinger solver remains the tool of choice.
Nonetheless,
the construction method described here, because of its speed and accuracy, holds great promise for several applications with direct observational signatures. These include the implications of the wave interference substructures for stellar heating \cite{Hui:2016ltb,Bar-Or:2018pxz,Church:2018sro,Marsh:2018zyw,El-Zant:2019ios,DuttaChowdhury2021}, the scattering of tidal streams \cite{Amorisco:2018dcn,Schutz:2020jox,Benito:2020avv,Dalal:2020mjw}, and gravitational lensing \cite{Chan:2020exg,Hui:2020hbq}.
We hope to address some of these issues in the future.

\acknowledgments

We wish to thank Kathryn Johnston, Jerry Ostriker, and Scott Tremaine for helpful discussions, and especially Neal Dalal for discussions regarding the relation between superposition amplitudes and the particle distribution function. We additionally wish to thank the anonymous referee for detailed and insightful comments that helped improve this article.

The dynamical simulations described in this work were performed using \texttt{Enzo}\footnote{http://enzo-project.org} \citep{ENZO:2013hhu} and analyzed with the \texttt{YT} toolkit \citep{yt_method}, both of which are publicly available. In addition, we have made extensive use of \texttt{Astropy} \citep{Astropy:2013muo,Price-Whelan:2018hus}, \texttt{Matplotlib} \citep{Hunter2007}, \texttt{Numpy} \citep{VanDerWalt2011}, \texttt{Scipy} \citep{Virtanen:2019joe}, and \texttt{SHTools} \citep{Wieczorek2018}. The dynamical simulations were run on Columbia University's \textit{Habanero} and \textit{Ginsburg} HPC clusters, and we acknowledge computing resources from Columbia University's Shared Research Computing Facility project, which is supported by NIH Research Facility Improvement Grant 1G20RR030893-01, and associated funds from the New York State Empire State Development, Division of Science Technology and Innovation (NYSTAR) Contract C090171, both awarded April 15, 2010.

TDY is supported through the NSF Graduate Research Fellowship (DGE-1644869). TDY is also partially supported by the National Science Foundation under Grant No. AST-1715582. XL is supported by the Natural Sciences and Engineering Research Council of Canada (NSERC), funding reference \#CITA 490888-16 and the Jeffrey L. Bishop Fellowship.
Research at Perimeter Institute is supported in part by the Government of Canada through the Department of Innovation, Science and Economic Development Canada and by the Province of Ontario through the Ministry of Colleges and Universities.  LH is supported by the DOE DE-SC0011941 and a Simons Fellowship in Theoretical Physics.

\appendix

\section{Connecting waves and particles -- the WKB limit} \label{sec:app}


In this Appendix, we wish to clarify the relation between the particle distribution function
and the wave superposition coefficients. 
The two main results are Eqs. (\ref{aiNe}) and (\ref{feai}).
Eq. (\ref{aiNe}) gives the general relation between $|a_j|^2$ and $f$ without using the WKB approximation (but it does assume a continuum limit such that the number of eigenstates per energy is well-defined). Eq. (\ref{feai}) provides the WKB limit of this relation.
Along the way, we review a few well known results from galactic dynamics,
and as a bonus, we will see how the wave description provides a 
convenient way to understand them.

We wish to relate three different expressions for the density $\rho$ in a halo.
The first is:
\begin{eqnarray}
\rho (\vec x) = \int d^3 v f(\vec x , \vec v) \, ,
\label{rhof}
\end{eqnarray}
where $f$ is the mass distribution function for particles (each of mass $m_a$), with $f$ telling us the amount of mass
per phase space volume ($\vec x$ and $\vec v$ are position and velocity). 
The second is: 
\begin{eqnarray}
\rho (\vec x) = \sum_\alpha M_\alpha P_\alpha (\vec x) \, .
\label{rhoP}
\end{eqnarray}
where $\alpha$ labels particle orbits, $M_\alpha$ is the amount of mass contained in particles that belong to orbit $\alpha$,
and $P_\alpha (\vec x) d^3 x$ is the probability that a particle of orbital type $\alpha$ happens
to be in the vicinity of $\vec x$ (within volume $d^3 x$, i.e., $P_\alpha (\vec x) d^3 x$ is
the fraction of time particles of orbit $\alpha$ spend in that
volume). The classic Schwarzschild method for constructing a halo is:
for a desired potential, find all possible orbits, and assign weights $M_\alpha$ such that $\rho$ matches the
corresponding desired density profile. 

The above two expressions for $\rho$ are appropriate for a halo composed of particles.
For a halo composed of waves, we have:
\begin{eqnarray}
\rho (\vec x) = \sum_j m_a |a_j|^2 |\psi_j (\vec x)|^2 + ...\, ,
\label{rhopsi}
\end{eqnarray}
where it is assumed the wave function $\Psi = \sum_j a_j \psi_j$ with
$j$ labeling the eigenmode $\psi_j$ and $a_j$ is its amplitude.
In squaring the wave function to obtain $\rho$, we will ignore interference terms
represented by the ellipsis, i.e., $\rho$ here can be thought of as
the time-averaged profile.
Our normalization convention is $\int d^3 x |\psi_j |^2 = 1$. 
There is a factor of particle mass $m_a$ because the amplitude $a_j$ is dimensionless.

To connect Eqs. (\ref{rhof}), (\ref{rhoP}) and (\ref{rhopsi}), it is
helpful to have a simple, concrete example in mind. 
Let us focus on a spherically symmetric halo, with a distribution
function $f$ that is a function of energy alone, or energy per unit
mass: $e \equiv E/m_a = |\vec v|^2/2 + V$, where $V$ is the
gravitational potential. Such a distribution $f(e)$ solves the
collisionless Boltzmann equation, and implies an isotropic local
velocity dispersion.
Eq. (\ref{rhof}) can be rewritten as:
\begin{eqnarray}
\label{rhofe}
\rho(r) = 4 \pi \int_{V(r)}^0 de f(e) \sqrt{2(e - V(r))} \, ,
\end{eqnarray}
where we have assumed a bound halo such that the maximum $e = 0$ \, .
Note that our $e$ is opposite in sign to
${\cal E}$ in \cite{BT}.
It is possible to invert the above equation, though we 
do not need it for the discussion in this Appendix:
assuming $\rho$ and $V$ are
monotonic functions of the radius $r$, we have upon differentiation:
\begin{eqnarray}
\frac{d\rho}{d V} = - {2 \sqrt{2} \pi} \int_V^0 de \frac{f(e)}{\sqrt{e - V}} \, ,
\end{eqnarray}
which can be inverted to give the Eddington formula:
\begin{eqnarray}
\label{eddington}
f(e) = \frac{1}{2 \sqrt{2} \pi^2} \frac{d}{de} \left( \int_e^0 dV \frac{d\rho/dV}{\sqrt{V - e}} \right) \, .
\end{eqnarray}
To show this, observe that 
$\int_e^0 dV {(d\rho/dV) /\sqrt{V - e}} = -2 \sqrt{2} \pi^2 \int_e^0 d\tilde e f(\tilde e)$, by
plugging in the expression for $d\rho/dV$, exchanging the order of integration
and using 
$\int_e^{\tilde e} dV /\sqrt{(V-e)(\tilde e - V)} = \int_0^1 dz/\sqrt{z (1-z)} = \pi$.

It is interesting to compare Eqs. (\ref{rhofe}) and (\ref{rhoP}). 
Essentially, the energy per unit mass $e$ plays the role of $\alpha$ in 
Eq. (\ref{rhoP}): all orbits with the same $e$ are given the same
weight $M_\alpha$. 
The factor $\sqrt{2(e - V(r))}$ must somehow be proportional to 
the probability $P_\alpha$ of finding a particle at radius $r$. 
It might come as a surprise that the probability should scale as
$\sqrt{2(e - V)}$, which is like a velocity -- after all, one would think the higher the velocity,
the less time the particle would spend at that location. This reasoning
turns out to miss an important effect. As we will see, the wave picture gives us a convenient
way (not the only way, of course) to understand this.

Let us start by comparing Eq. (\ref{rhofe}) with Eq. (\ref{rhopsi}).
Integrating Eq. (\ref{rhopsi})  over volume, we have:
\begin{eqnarray}
\label{rhoNe}
\int d^3 x \rho = \sum_j m_a |a_j|^2 \sim \int de  \, N(e) \, m_a |a_j|^2 \, ,
\end{eqnarray}
where we assume $|a_j|^2$ is a function of $e$ only (i.e., a wave analog of $f(e)$), and 
$de N(e)$ represents the number of states around $e \pm de/2$. 
On the other hand, integrating Eq. (\ref{rhofe}) over volume gives,
upon exchanging the order of integration and assuming spherical symmetry:
\begin{equation}
\int d^3x \, \rho = 
4\pi \int_{V(0)}^0 de f(e) \int_0^{\rm r_{\rm max} (e)} dr 4\pi r^2
\sqrt{2 (e - V(r))} \, ,
\end{equation}
where $r_{\rm max} (e)$ is the maximal radius reached by a particle of energy $E = m_a e$.
Comparing the two expressions, it is natural to equate:
\vspace{.15cm}
\begin{tcolorbox}[colframe=white,arc=0pt,colback=greyish2]
\vspace{-.42cm}
\begin{equation}
\label{aiNe}
|a_j|^2 = f(e) \frac{4 \pi}{m_a N(e)} \int_0^{\rm r_{\rm max} (e)} dr 4 \pi r^2 \sqrt{2 (e - V(r))} \, .
\end{equation}
\end{tcolorbox}
We wish to show, in the WKB limit, this simplifies to $|a_j|^2 \sim
f(e)$ times a constant factor. 
And as a bonus, we will understand better why in the integral for
$\rho$, Eq. (\ref{rhofe}), orbits with a seemingly higher velocity are
given a higher weight.

Consider the eigenmode $\psi_j$ (substituting for the role of an ``orbit''),
written as $R_j (r) Y^{m}_\ell (\theta,\phi)$ with the radial function $R_j$ satisfying:
\begin{equation}
- \frac{\hbar^2}{2m_a}  \partial_r^2 (r R_j) + \left( \frac{\hbar^2}{2 m_a}\frac{\ell(\ell+1)}{r^2} + m_a V \right) r R_j = E r R_j \, .
\end{equation}
Here, the label $j$ stands for $n, \ell, m$ (the radial, angular, and magnetic quantum numbers). With a spherically symmetric $V$, the energy $E$ and
the radial function $R_j$ depend on $n$ and $\ell$
but not $m$. 
Using the fact that $\sum_m |Y_\ell^m |^2 = (2\ell + 1)/(4\pi)$, and assuming
$a_i$ is independent of $m$ (as is appropriate for the wave
analog of a halo with a distribution function $f(e)$), we have:
\begin{eqnarray}
\rho(r) = \frac{m_a}{4\pi} \sum_{n,\ell} (2\ell + 1) |a_{n\ell}|^2 |R_{n\ell} (r)|^2 \, .
\end{eqnarray}
For the radial function, we can use the WKB approximation in the large
$E$ limit \cite{Scrucca2012}:
\begin{equation}
\label{Rnl}
R_{n\ell} (r) = \frac{{\cal N}_{n\ell}}{r \sqrt{k_{\rm eff} (r)}}
{\,\rm sin\,} \left[ \int_{r_1}^r k_{\rm eff} (r') dr'/\hbar + \frac{\pi}{4} \right] \, ,
\end{equation}
where:
\begin{eqnarray}
k_{\rm eff} (r) \equiv \sqrt{ 2m_a \left(E - \frac{\ell(\ell+1)}{2 m_a r^2} - m_a V\right)} \nonumber \\
= m_a \sqrt{ 2 \left(e - \frac{\ell(\ell+1) \hbar^2}{2 m_a^2 r^2} - V\right)} \, .
\end{eqnarray}
and the energy eigenvalue $E$ satisfies the quantization condition:
\begin{eqnarray}
\label{pin}
\pi \left(n + \frac{1}{2}\right) = \int_{r_1}^{r_2} k_{\rm eff} (r) dr/\hbar \, .
\end{eqnarray}
Here, $r_1$ and $r_2$ are the turning points where $k_{\rm eff}$ vanishes.
The normalization ${\cal N}_{n\ell}$ is chosen such that
$\int d^3 x |\psi_j (\vec x)|^2 = 1$:
\begin{eqnarray}
{\cal N}_{n\ell}^2 = 
\left[ \int_{r_1}^{r_2} \frac{dr}{2 k_{\rm eff} (r)} \right]^{-1} \, ,
\end{eqnarray}
where we have approximated the square of the sine as $1/2$
and used $\int {\,\rm sin}\theta \, d\theta d\phi |Y^m_\ell |^2 = 1$.
We approximate the sum over $n$ and $\ell$
by integrals:
\begin{eqnarray}
\label{rhodnde}
\rho(r) = \frac{m_a}{4\pi} \int de \, d\ell \frac{dn}{de} (2\ell + 1) |a_{\rm n\ell}|^2 |R_{n\ell} (r)|^2
\, .
\end{eqnarray}
The integration measure $dn/de$ can be obtained by differentiating
the quantization condition:
\begin{eqnarray}
\frac{dn}{de} = \frac{m_a^2}{\pi} \int_{r_1}^{r_2} \frac{dr}{k_{\rm
  eff} (r) \hbar} \, .
\end{eqnarray}
When differentiating the quantization condition, there are in principle contributions from the fact that $r_1$ and $r_2$ depend on $e$, but they turn out to vanish because the integrand $k_{\rm eff}$ vanishes precisely at these points.
Using this expression together with the WKB approximation
for $R_{n\ell}$ in Eq. (\ref{rhodnde}), and integrating over $\ell$,
we find:\footnote{For a given $e$ and $r$, $\ell$ ranges from $0$ to $\ell_{\rm max}$ such that
$e - [\ell_{\rm max}(\ell_{\rm max} + 1)\hbar^2/(2 m_a^2 r^2)] - V(r)
= 0$. The integration measure $(2\ell + 1)d\ell$ can be recast as $d(\ell[\ell+1])$. 
The integral turns out to be dominated by small rather than large
$\ell$; there is thus the concern that an integral approximation
of the sum over $\ell$ might not be accurate.
The point is that as long as $e$ is sufficiently large compared to $V(0)$,
the error made is small. 
}
\begin{eqnarray}
\rho(r) = \frac{m_a^4}{2 \pi^2 \hbar^3} \int de |a_j|^2 \sqrt{2 (e -V)} \, .
\end{eqnarray}
Here, it is important to be clear about the notation: the $j$ of $a_j$
still labels $n, \ell, m$, but $a_j$ is a function of $e$
only (wave analog of $f(e)$). We can compare this against the particle
distribution function description in Eq. (\ref{rhofe}). 
The two expressions match up nicely provided we identify:
\vspace{.15cm}
\begin{tcolorbox}[colframe=white,arc=0pt,colback=greyish2]
\vspace{-.42cm}
\begin{eqnarray}
\label{feai}
|a_j|^2 \sim \frac{(2\pi \hbar)^3}{m_a^4}  f(e) \, .
\end{eqnarray}
\end{tcolorbox}
The superposition coefficient $a_j$ is dimensionless, as it should be
(recall that $f(e) d^3x d^3 v$ has the dimension of mass).
We emphasize that the equality is approximate, in the sense it is derived in the WKB limit.
In the wave construction of a halo of a given density profile $\rho(r)$, we expect
the above to hold for eigenmodes with a high energy $e \gg V(0)$, i.e., modes that
have a significant overlap with the outer parts of the halo.
Incidentally, one could also derive Eq. (\ref{feai}) from Eq. (\ref{aiNe}), by
noting that $N(e) \sim \int d\ell \, (2\ell+1) (dn/de) \sim [2 m_a^3/(\pi \hbar^3)] \int_{0}^{r_{\rm max} (e)} r^2 dr \sqrt{2(e - V)}$. 

As a bonus, the above derivation shows why the contribution of
a given mode to the density at $r$ scales as $\sqrt{2(e - V)}$
(recall the puzzle stated earlier in the particle picture: that
according to
Eq. (\ref{rhofe}), the
probability density $P_\alpha$ scales as $\sqrt{2(e-V)}$, some sort of velocity).
From the form of the radial function in Eq. (\ref{Rnl}), 
we see that the contribution scales as $|R_{n\ell}|^2 \propto 1/k_{\rm
  eff}(r)$, 
which is as it should be: the larger the momentum or velocity, the less time a particle spends
at the location of interest $r$ (or the smaller the amplitude of the
wave mode at that position). 
What counteracts this is the integral over $\ell$: there are many 
more $\ell$ modes if $\sqrt{2(e-V)}$ is large.

To summarize, we have shown that for a spherically symmetric halo with 
a particle distribution function $f(e)$, the density profile is given
by Eq. (\ref{rhofe}) and the wave analog is Eq. (\ref{rhopsi}), with 
$|a_j|^2$ given by Eq. (\ref{feai}) in the WKB, i.e., high energy, limit.


\begin{figure}[ht!]
\includegraphics[width=\columnwidth]{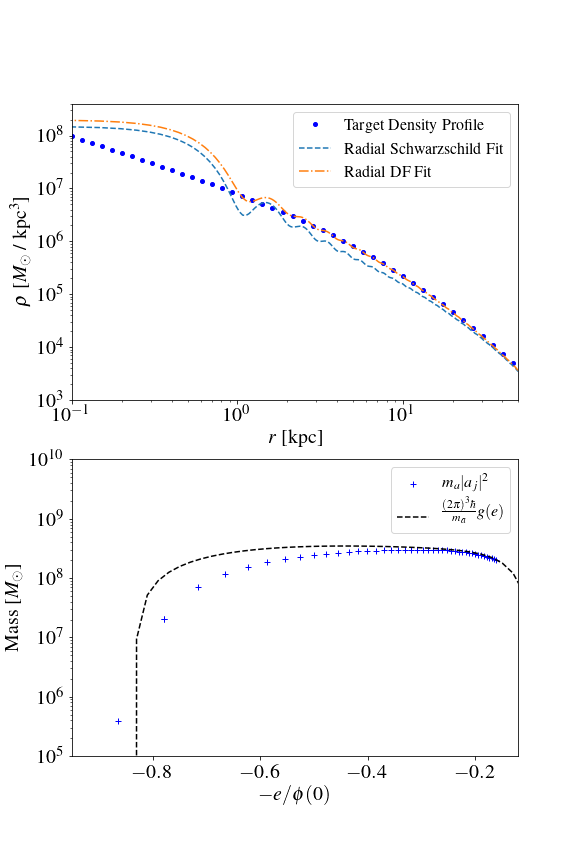} 
\caption{Top panel: Comparison of a target NFW density profile (blue dots) to the output density profiles of two constructed halos. The light blue dashed line represents the profile of a halo constructed using the Schwarzschild construction method, with a radial constraint applied (i.e., the amplitudes of all 
$\ell > 0$ modes are set to zero). The orange dot-dashed line represents the profile of a halo constructed directly from the radial distribution function (Eqs. \ref{eq:rad_fE} and \ref{aig}). The latter fits almost perfectly at large $r$ but deviates significantly closer to $r=0$. The Schwarzschild fit sacrifices the fit at large $r$ in order to reduce the deviation at small $r$. Bottom panel: Comparison of the scaled radial distribution function obtained through the Eddington inversion formula (dashed black line) to the squared amplitudes obtained through the Schwarzschild construction method.
}
\label{fig:app_1}
\end{figure}

Let us close with another simple example, a halo constructed entirely
out of $\ell = 0$ modes (i.e., the wave analog of a halo composed of
particles on radial orbits). 
Suppose $f = g(e) \delta_D (L^2)$ where $L$ is the angular momentum
per unit mass and $\delta_D$ is the Dirac delta function, the density profile is:
\begin{eqnarray}
\label{rhog}
&& \rho(r) = \int d^3 v \, g(e) \, \delta_D (r^2 v_\perp^2)
= \frac{2\pi}{r^2} \int_0^\infty dv_r g(e) \nonumber\\ 
&& \quad \quad = \frac{\sqrt{2} \pi}{r^2} \int_{V(r)}^0 de
  \frac{g(e)}{\sqrt{e - V(r)}} \, ,
\end{eqnarray}
where we use $e = v_r^2/2 + V(r)$. 
Note how the integrand is smaller at radii where the velocity is higher, in accord with expectation.
Again, we do not need it for our
discussion here, but for completeness, the inversion formula is
(viewing $\rho$ as a function of $V$):
\begin{eqnarray} \label{eq:rad_fE}
g(e) = - \frac{1}{\sqrt{2} \pi^2} \frac{d}{de} 
\left( \int_e^0 dV  \frac{\rho r^2}{\sqrt{V-e}} \right) \, ,
\end{eqnarray}
using the fact that
$\int_e^0 dV r^2\rho / \sqrt{V-e} = \sqrt{2} \pi^2 \int_e^0 d\tilde e
g(\tilde e)$, which can be verified by substituting in the expression for $\rho$. 

The volume integral of $\rho$ is:
\begin{eqnarray}
&& \int_0^\infty dr \, 4\pi r^2 \rho(r) =  {4 \sqrt{2} \pi^2} \int_0^\infty dr \, \int_{V(r)}^0 de
  \frac{g(e)}{\sqrt{e - V(r)}} \nonumber \\
&& \quad = {4 \sqrt{2} \pi^2} \int_{V(0)}^0 de \int_0^{r_{\rm max}(e)} dr
  \frac{g(e)}{\sqrt{e - V(r)}}  \, .
\end{eqnarray}
Comparing this against Eq. (\ref{rhoNe}), it is natural to equate:
\begin{eqnarray}
\label{ajradial}
|a_j|^2 = g(e) \frac{4 \sqrt{2} \pi^2}{m_a N(e) } \int_0^{r_{\rm max}(e)} 
  \frac{dr}{\sqrt{e - V(r)}} \, .
\end{eqnarray}

To make further progress, we use the WKB approximation.
The eigenmode $\psi_j (r)$ obeys:
\begin{eqnarray}
- \frac{\hbar^2}{2m_a}  \partial_r^2 (r \psi_j) + m_a V  r \psi_j = E r \psi_j \, ,
\end{eqnarray}
with the approximate solution:
\begin{equation}
\label{psii}
\psi_j (r) = \frac{\cal N}{r \sqrt{k_{\rm eff} (r)}}
{\,\rm sin\,} \left[ \int_{0}^r k_{\rm eff} (r') dr'/\hbar + \frac{\pi}{4} \right] \, ,
\end{equation}
where:
\begin{equation}
k_{\rm eff} (r) \equiv \sqrt{ 2m_a \left(E - m_a V\right)}
= m_a \sqrt{ 2 \left(e - V\right)} \, ,
\end{equation}
and the quantization condition \cite{weinberg2015}:
\begin{eqnarray}
\label{pin0}
\pi \left( n + \frac{3}{4} \right) = \int_{0}^{r_2} k_{\rm eff} (r)
  dr/\hbar \, ,
\end{eqnarray}
where $r_2$ is the outer turn-around radius. Here, with no $\ell$ and
$m$, the label $j$ is the same as $n$. 
The normalization ${\cal N}$ is chosen to keep
$\int dr 4\pi r^2 |\psi_j|^2 = 1$:
\begin{eqnarray}
{\cal N}^2 = 
\left[ \int_{0}^{r_2} \frac{2\pi dr}{k_{\rm eff} (r)} \right]^{-1} \, .
\end{eqnarray}
Differentiating the quantization condition, we have:
\begin{eqnarray}
\label{dnde2}
\frac{dn}{de} = \frac{m_a^2}{\pi} \int_{0}^{r_2} \frac{dr}{k_{\rm eff}
  (r) \hbar} \, .
\end{eqnarray}
To find $|a_j|^2$, we can follow one of two approaches.
One is to use Eq. (\ref{ajradial}), with 
$N(e) \sim dn/de$, giving:
\begin{eqnarray}
\label{aig}
|a_j|^2 = \frac{(2\pi)^3 \hbar}{m_a^2} g(e) \, .
\end{eqnarray}
The other is to equate Eq. (\ref{rhog}) with $\rho(r) = \sum_j m_a |a_j|^2
|\psi_j (r)|^2 \sim \int de (dn/de) m_a |a_j|^2 |\psi_j (r)|^2$, making
use of Eqs. (\ref{dnde2}) and (\ref{psii}). One confirms
Eq. (\ref{aig}), and sees that the WKB prefactor of
$|\psi_j|^2$ nicely reproduces $1/\sqrt{e-V}$
in the integrand of Eq. (\ref{rhog}).

Figure \ref{fig:app_1} compares the output density profiles and the eigenmode amplitudes of two halos constructed to match the NFW target density profile used throughout the paper (with no core; see Eq. \ref{eq:NFW}). The first is constructed using the Schwarzschild method, constrained to only allow radial modes; the amplitudes of all $\ell>0$ modes are set to zero. The second is constructed directly from $g(e)$ following Eqs. (\ref{eq:rad_fE}) and (\ref{aig}). The fit from $g(e)$ matches the outer profile (dominated by eigenmodes at the higher energies for which the WKB approximation holds) almost perfectly, and deviates from the target only at smaller radii. The Schwarzschild fit is similarly good, though the algorithm has sacrificed the fit at large radii in order to reduce the deviation from the target profile at smaller radii. In the high energy limit, both fits assign almost identical amplitudes to the eigenmodes, as shown in the lower panel of Figure \ref{fig:app_1}.

\bibliography{references}

\end{document}